\documentclass[sigconf, nonacm]{acmart}

\newcommand\vldbavailabilityurl{}

\usepackage{graphicx}
\usepackage{balance}  
\usepackage{url}
\usepackage{xcolor}
\usepackage{listings}
\usepackage[a-1b]{pdfx}

\lstnewenvironment{DSL}
  {\lstset{
        aboveskip=5pt,
        belowskip=5pt,
        escapechar=!,
        mathescape=true,
        language=SQL,
        basicstyle=\linespread{0.96}\ttfamily\small,
        morekeywords={
        Map, HashMap, aggregate, join, select, integer
        },
        deletekeywords={VALUE, PRIOR},
        showstringspaces=true}
        \vspace{0pt}%
        \noindent\minipage{0.47\textwidth}}
  {\endminipage\vspace{0pt}}
  
\newcommand{\concat}{{\textsc{Concat}}}
\newcommand{\rekey}{{\textsc{ReKey}}}
\newcommand{\tile}{{\textsc{Tile}}}

\newcommand{\shuffle}{{\textsc{Shuf}}}
\newcommand{\broadcast}{{\textsc{Bcast}}}
\newcommand{\localjoin}{{\textsc{$\Join^{L}$}}}
\newcommand{\localagg}{{\textsc{$\Sigma^{L}$}}}
\newcommand{\localmap}{{\textsc{$\lambda^{L}$}}}
\newcommand{\localfilter}{{\textsc{$\sigma^{L}$}}}

\newcommand{\frontier}{{\textsc{Front}}}

\newcommand{\phys}[1]{\mathsf{#1}}

\begin{document}
\title{Tensor Relational Algebra for Distributed Machine Learning System Design}

\author{Binhang Yuan$^\dagger$, Dimitrije Jankov$^\dagger$, Jia Zou$^\ddagger$, Yuxin Tang$^\dagger$, Daniel Bourgeois$^\dagger$, Chris Jermaine$^\dagger$}
\affiliation{
 \institution{
    \text{$^\dagger$Rice University; $^\ddagger$Arizona State University}\\
    \text{$^\dagger$\{by8, dj16, yuxin.tang, dcb10, cmj4\}@rice.edu; $^\ddagger$jia.zou@asu.edu}}
}

\begin{abstract}
We consider the question: what is the  abstraction that should be implemented by the computational engine of a machine learning system? Current machine learning systems  typically push whole tensors through a series of compute kernels such as matrix multiplications or activation functions, where each kernel runs on an AI accelerator (ASIC) such as a GPU.  This implementation abstraction provides little built-in support for ML systems to scale past a single machine, or for handling large models with matrices or tensors that do not easily fit into the RAM of an ASIC. In this paper, we present an alternative implementation abstraction called the \textit{tensor relational algebra} (TRA). The TRA is a set-based algebra based on the relational algebra. Expressions in the TRA operate over binary tensor relations, where keys are multi-dimensional arrays and values are tensors. The TRA is easily executed with high efficiency in a parallel or distributed environment, and amenable to automatic optimization. 
Our empirical study shows that the optimized TRA-based back-end can significantly outperform alternatives for running ML workflows in distributed clusters.
\end{abstract}

\maketitle

\ifdefempty{\vldbavailabilityurl}{}{
\vspace{.3cm}
\begingroup\small\noindent\raggedright\textbf{PVLDB Artifact Availability:}\\
The source code, data, and/or other artifacts have been made available at \url{\vldbavailabilityurl}.
\endgroup
}

\section{Introduction}

In widely-used machine learning (ML) systems such as TensorFlow~\cite{tensorflow} and PyTorch~\cite{pytorch}, computations are usually specified operationally: a sequence of mathematical operations such as matrix multiplications, activation functions (ReLU, tanh), convolutions, etc., are applied to a set of input tensors to define a computation. HPC systems such as ScaLAPACK~\cite{choi1992scalapack} and distributed analytic packages such as Dask~\cite{dask} offer a similar, operational interface.  Operations are ``black boxes'' in the sense that the internals of the operation are mostly opaque to the system. An operation such as a matrix multiply is not a logical operation that the ML system figures out how to best optimize at runtime. Instead, it is a physical operation that has to run somewhere, on some hardware, via the invocation of a computational kernel. 

This operational approach has certain drawbacks, namely that the system has limited latitude for automatic optimization. The programmer is responsible for making sure that the operations can run successfully using available resources (such as the amount of RAM on each GPU), and if the operations cannot run successfully, the programmer must figure out how to break the operations up into smaller pieces  that can run. Tasks such as getting a computation to run on multiple GPUs or on multiple machines in a distributed cluster so as to minimize communication are left to the programmer.

\vspace{5 pt}
\noindent
\textbf{Toward declarative tensor programming.}
There has been work on programming models that are more declarative. PyTorch and TensorFlow now both support variants of Einstein notation---a classical notation for specifying operations over tensors, and work on optimizing such computations has made its way into commercial systems~\cite{abo2016faq}. Researchers have proposed variants of the Ricci calculus as a specification language for ML~\cite{laue2020simple}. There have been other proposals for declarative tensor programming languages that allow for automatic generation of compute kernels that can be optimized to handle the data at hand such as Tensor Comprehensions~\cite{vasilache2018tensor}.

Nevertheless, while there has been attention paid at the question of how to specify ML computations declaratively, there has been little attention paid to the question of what the correct implementation abstraction for ML system design should be.  That is, what target should a ML system \emph{back-end} present to the \emph{front-end}?\footnote{Throughout the paper, we use the term \emph{front-end} to refer to the programmer-facing API and compiler; in PyTorch, for example, this would be the part of the system that accepts Einstein notation and transforms it into a set of executable operations.  We use the term \emph{back-end} to refer to the sub-system that actually executes the computation.} There are several requirements for such a back-end interface.  It should be able to express most/all important ML or numerical computations. It should be hardware agnostic, but computations specified using the interface should be easily scaled to multiple ASICs and multiple machines.  It should allow for computations over very large input data.  It should facilitate easy, automatic optimization.  And it should provide for execution times that are as fast as a carefully-designed, ``hand-built'' computation on top of the very best tools such as ScaLAPACK, Dask, TensorFlow, or PyTorch.  

\vspace{5 pt}
\noindent
\textbf{The tensor relational algebra.}
In this paper, we argue that the implementation abstraction that should be offered by a ML system back-end is the \emph{tensor relational algebra}, or TRA for short. The TRA is a simple variant of the classical relational algebra (RA), which serves as the standard implementation abstraction for modern relational database system design. The key difference is that in the TRA, the relations over which computations are performed are always binary relations between $k$-dimensional vectors (keys) and $r$-dimensional tensors.

Of course, it is widely understood that a $k$-dimensional tensor can be stored as a binary relation between $k$-dimensional keys and real number values, e.g.~\cite{hutchison2017laradb}. Thus, why not use classical RA as the implementation abstraction?  There is good evidence that a compute engine based upon such an abstraction will perform very poorly over the dense-tensor computations common in deep learning~\cite{luo2018scalable}: the overhead associated with storing each entry in a high-dimensional tensor as a separate tuple and moving billions of tuples through a system can result in poor performance, especially when the competition is a high-performance CPU or GPU kernel.  This is why the TRA specifically allows for tensors or ``chunks'' of a larger tensor to be stored in each tuple---the fixed, per-tuple cost is incurred by a much smaller number of tuples. 

We argue that the TRA has three key advantages as an implementation abstraction: expressivity, easy optimization, and high performance. The joins and aggregations offered by the TRA can implement the indexed operations required by the Einstein notation and related specification languages. It is easy to implement and optimize relational operations in a distributed environment, as the decades of success enjoyed by relational database systems has demonstrated. And by design, an ML system back-end implementing the TRA is able to run classical HPC algorithms which rely on decomposing matrices into chunks, and it can run them almost as well as special-purpose HPC softwares such as ScaLAPACK.  

\vspace{5 pt}
\noindent
\textbf{Our Contributions.}  
We propose the TRA as well as an implementation algebra (IA) which is easy to implement in a distributed system.  We consider how computations in the TRA can be transformed into computations in the IA, and propose a set of transformations or equivalences that allow re-writes of computations in the IA.  Finally, we implement a prototype of the IA, and show that it can enable efficient, distributed implementations of ML computations, which can reach or even significantly outperform the existing HPC and ML systems including ScaLAPACK, Dask, TensorFlow, and PyTorch.

\section{TRA Overview}

\subsection{Motivation}

There has been significant  interest in tensor manipulation languages, which allow for declarative specification of ML and numerical computations. This simplest of these is the Einstein notation, which provides a way to write a tensor computation like the form: 
$$\forall i,j: \textrm{ }\textbf{C}_{i,j} \leftarrow \sum_k \textbf{A}_{i,k} \times \textbf{B}_{k,j}.$$
\noindent This example describes matrix multiplication. 
The value $i$-th row and $j$-th column of the output is the dot product of the $i$-th row of input matrix $\textbf{A}$ and the $j$-th column of input matrix $\textbf{B}$.  
Different proposals have different variations on this idea~\cite{abo2016faq,laue2020simple,vasilache2018tensor}, but the common features are that (1) values from different tensors are fed into a scalar function (such as the multiplication above) by matching indices, and (2) 
those dimensions are summed over.

Languages such as the Einstein notation provide an excellent, declarative interface for programming an ML system---much as SQL provides the interface for relational systems. But the question remains: what is the correct implementation abstraction for ML systems?  That is, what is the interface that the back-end should export, which can be targeted by an ML programming system compiler and auto-differentiation engine that make up the ML system's front-end? 

\subsection{TRA: The Basics}

We propose the TRA as this ML implementation abstraction.
The TRA operates over \emph{tensor relations} containing pairs of the form: 
$$\texttt{(key, array)}.$$
Conceptually, these tensor relations store sets of arrays.
Each key value serves, not surprisingly, as the key for the pair.

Tensors are decomposed into sets of sub-tensors to represent them as tensor relations. For example, consider the matrix \textbf{A}, 
$$\textbf{A} = \begin{bmatrix}
1 & 2 & 5 & 6 \\
3 & 4 & 7 & 8 \\
9 & 10 & 13 & 14 \\
11 & 12 & 15 & 16
\end{bmatrix},$$
we may store this as a tensor relation
\begin{align} \texttt{R}_A = \Bigg\{& \left( \langle 0, 0 \rangle, \begin{bmatrix} 1 & 2 \\ 3 & 4 \end{bmatrix}  \right),
        \left( \langle 0, 1 \rangle, \begin{bmatrix} 5 & 6 \\ 7 & 8 \end{bmatrix}  \right), \nonumber \\
        &\left( \langle 1, 0 \rangle, \begin{bmatrix} 9 & 10 \\ 11 & 12 \end{bmatrix}  \right),
        \left( \langle 1, 1 \rangle, \begin{bmatrix} 13 & 14 \\ 15 & 16 \end{bmatrix}  \right) \nonumber
\Bigg\}.\end{align}

The TRA offers a similar set of operations to the RA: joins, aggregations, selections. It makes an excellent compilation target for tensor manipulation languages like the Einstein notation for several reasons. First, these languages typically match elements in tensors based on shared index values (which can be implemented as relational joins) and then sum out various dimensions (implemented as aggregations).    
The problem with using the RA as an implementation abstraction for tensors, where tensors are stored relationally as keys identifying a single non-zero value in a tensor, is performance. Tensors can have millions or billions of entries, and it seems impossible to build a back-end with acceptable performance if it has to process one tuple per non-zero value.  

Thus, the TRA operates over sub-tensors or ``chunks'', and expects the ML system front-end to supply a kernel function (typically a high-performance CPU or GPU operation) to operate over the sub-tensors themselves.  This does complicate the TRA compared to the RA, but as we show experimentally, this modification makes it possible for a TRA-based back-end to significantly outperform the back-ends offered by TensorFlow and PyTorch.

\subsection{TRA: the Implementation Algebra}

One of the key advantages of the TRA is that like the RA, there are a number of re-writes, inherited from the RA (such as push-down of selections) that can be used to optimize TRA computations.

However, when designing an ML system back-end, one of our key goals is to distribute ML computations across multiple ASICs or multiple machines.  Such distributed computations have many different implementations.  Consider the matrix multiply required to push a batch of feature vectors (represented as one matrix) through the links into a fully connected layer in a neural network (the weights are represented as a second matrix). This could be implemented by decomposing the feature matrix into sub-matrices at each site (each containing a subset of the feature vectors) and broadcasting the weight matrix to all sites. This is the common ``data parallel'' implementation, in ML system parlance. Or, one could fully decompose both matrices and apply a complex distributed algorithm, such as the 3D algorithm~\cite{agarwal1995three}. This would be a combined data and ``model parallel'' implementation in ML system parlance.  Crucially, the TRA cannot express the distinctions among such implementations.

As such, we also propose an \emph{implementation algebra} (IA) that can express these different distributed computations, as well as a simple compilation strategy from the TRA to the IA.  Further, we propose an extensive set of re-writes for the IA that allow such disparate implementations to be reached from an initial TRA computation, and a simple cost model.  Given this, an ML system back-end would export a simple, TRA-based interface which says nothing about distribution to multiple machines or ASICs.  A TRA computation can then be compiled into a computation in the IA, and optimized into a high-performance, distributed computation.

\section{Tensor Relational Algebra}

The RA operates over \texttt{(key, array)} pairs. Define an \emph{array type} that consists of:

\begin{enumerate}
    \item 
A \emph{rank} $r \in \mathbb{Z}^{*}$ 
    \item
A \emph{bound} $\textbf{b} \in (\mathbb{Z}^{*})^r$.  
\end{enumerate}

\noindent For two vectors $\textbf{u} = \langle u_i \rangle$ and $\textbf{v} = \langle v_i \rangle$, define $\textbf{u} \leq \textbf{v} \equiv \wedge_i \left( u_i \leq v_i \right)$.
Define $\textbf{u} < \textbf{v}$ similarly.  
Informally, we say that an array of rank $r$ is \emph{bounded} by vector \textbf{b} if the array is $r$ dimensional, and for any $r$-dimensional vector \textbf{i} that is less than the bound, $\texttt{array}_{\textbf{i}}$ returns a real number. Formally:

\begin{enumerate}
    \item For any index  $\textbf{i} \in (\mathbb{Z}^*)^r$, $\vec{0} \leq \textbf{i} < \textbf{b} \implies \texttt{array}_{\textbf{i}} \in \mathbb{R}$.  
    \item $\neg (\vec{0} \leq \textbf{i} < \textbf{b}) \implies \texttt{array}_{\textbf{i}} = \bot$.  That is, for any index $\textbf{i}$ outside of the bound $[\vec{0}, \textbf{b}]$, $\texttt{array}_{\textbf{i}}$ is undefined.
\end{enumerate}

Subsequently, we denote the set of all arrays of rank $r$ and bound $\textbf{b}$ as $T^{(r, \textbf{b})}$. Thus, $T^{(r, \textbf{b})}$ defines an array type.  

We denote the power set of $(\mathbb{Z}^{*})^k \times T^{(r, \textbf{b})}$ as $R^{(k, r, \textbf{b})}$; this is the set of all possible tensor relations with $k$-dimensional keys, storing arrays of type $T^{(r, \textbf{b})}$.

\subsection{Operations in Tensor Relational Algebra}
Given this, the TRA is essentially a set of higher-order functions over tensor relations.  That is, each operation takes as input a kernel function defined over multi-dimensional arrays (in practice, this function is likely be an array-based MKL, CUDA, or Verilog kernel) and returns a function over tensor relations.  

We begin by giving an overview of the higher-order functions taking binary functions as input: aggregation (denoted using $\Sigma$) and join (denoted using $\Join$).

\noindent
(1) \textit{Aggregation} is a function:
\begin{align}
\Sigma : \left( (\mathbb{Z}^{*})^g \times \left(T^{(r, \textbf{b})} \times T^{(r, \textbf{b})}
\rightarrow T^{(r, \textbf{b})}\right)\right) \nonumber \\ 
\rightarrow \left(R^{(k, r, \textbf{b})}  \rightarrow R^{(g, r, \textbf{b})} \right) \nonumber 
\end{align}
\noindent$\Sigma _{\texttt{(groupByKeys, aggOp)}}$ takes as input a list of key dimensions to aggregate according to \texttt{groupByKeys} as well as an array kernel operation \texttt{aggOp}, and then returns a function that takes as input a tensor relation, groups the arrays in the relation based upon the indicated key values, and applies \texttt{aggOp} to the arrays in the group.

Consider the matrix $\textbf{A}$ from the last section. We can sum up the individual arrays vertically using $$\Sigma_{(\langle 1 \rangle, \texttt{matAdd})}(\texttt{R}_A)$$ \noindent which gives:
$$\left\{ \left( \langle 0 \rangle, \begin{bmatrix} 10 & 12 \\ 14 & 16 \end{bmatrix}  \right),\left( \langle 1 \rangle, \begin{bmatrix} 18 & 20 \\ 22 & 24 \end{bmatrix}\right)\right\}.$$
Because of the argument $\langle 1 \rangle$, the call $\Sigma _{(\langle 1 \rangle, \texttt{matAdd})}$ constructs an aggregation
function that groups all pairs having the same value for the key in position $1$, and sums them.
Or we could sum up the individual arrays into a single array using:
$$\Sigma _{(\langle \rangle, \texttt{matAdd})}(\texttt{R}_A)$$ which gives:
$$\left\{ \left( \langle \rangle, \begin{bmatrix} 28 & 32 \\ 36 & 40 \end{bmatrix}  \right) \right\}.$$

\noindent
(2) \textit{Join} is a function:
\begin{align} 
\Join : \bigg(& (\mathbb{Z}^{*})^{g} \times (\mathbb{Z}^{*})^{g} \times \left(T^{(r_l, \textbf{b}_l)} \times T^{(r_r, \textbf{b}_r)} \rightarrow T^{(r_o, \textbf{b}_o)}\right)\bigg) \nonumber \\
&\rightarrow \left(R^{(k_l, r_l, \textbf{b}_l)} \times R^{(k_r, r_r, \textbf{b}_r)} \rightarrow R^{(k_l + k_r - g, r_o, \textbf{b}_o)} \right) \nonumber 
\end{align}
\noindent $\Join _{\texttt{(joinKeysL, joinKeysR, projOp)}}$ takes as input a set of key dimensions to join on from the left and from the right, as well as an operation to run over all \texttt{(leftArray, rightArray)} pairs that are created during the join, and returns a function that performs the join and applies \texttt{projOp} to all pairs. Similar to a natural join in classical databases systems, the output key is all of the key values from the left input, with all of the key values from the right input appended to them, subject to the constraint that no value in \texttt{joinKeysR} is repeated a second time.

With join and aggregation we may implement matrix multiply over two matrices stored as tensor relations.
Imagine that we want to implement $\textbf{A} \times \textbf{A}$ for the matrix \textbf{A} defined
previously, where \textbf{A} is stored as a tensor relation $\texttt{R}_A$.  This can be written as:
$$\Sigma _{\left(\langle 0, 2 \rangle, \texttt{matAdd}\right)} \left(\Join _{\left(\langle 1 \rangle, \langle 0 \rangle, \texttt{matMul}\right)} \left(  \texttt{R}_A, \texttt{R}_A \right) \right)$$

\noindent This computes a matrix multiply of the matrix \textbf{A} because all of the pairs in $\texttt{R}_A$ are first joined on key index $1$ from the first instance of $\texttt{R}_A$ equaling key index $0$ from the second instance of $\texttt{R}_A$.
Each pair of arrays are then multiplied using the kernel $\texttt{matMul}$.
For example,
$$\left( \langle 0, 1 \rangle, \begin{bmatrix} 5 & 6 \\ 7 & 8 \end{bmatrix}  \right) \textrm{ and }
\left( \langle 1, 0 \rangle, \begin{bmatrix} 9 & 10 \\ 11 & 12 \end{bmatrix}  \right)$$ 
\noindent are joined to produce
$$\left( \langle 0, 1, 0 \rangle, \begin{bmatrix} 111 & 122 \\ 151 & 166 \end{bmatrix}  \right).$$  
\noindent The index $\langle 0, 1, 0 \rangle$ in this output pair
is a combination of
$\langle 0, 1 \rangle$ and $ \langle 1, 0 \rangle$ from the two input pairs, with the redundant index entry dropped (redundant because we know that two of the entries in positions 1 and 0, respectively, are repeated due to the join).  
Next, the arrays are aggregated using $\texttt{matAdd}$, summing out index $1$ (keeping indices $\langle 0, 2 \rangle$ as \texttt{groupByKeys}), to complete the matrix multiply.

In contrast to join and aggregation, rekey, filter and transform are higher-order functions taking a unary function as input.

\noindent
(3) \textit{ReKey} allows manipulation of keys: 
\begin{align}
\rekey: \left((\mathbb{Z}^{*})^{k_i} \rightarrow (\mathbb{Z}^{*})^{k_o}  \right) \rightarrow \left(R^{(k_i, r, \textbf{b})} \rightarrow R^{(k_o, r, \textbf{b})} \right) \nonumber \end{align}
\noindent $\rekey_{\texttt{(keyFunc)}}$ applies the $\texttt{keyFunc}$ on every key in the relation and generates a new key.

\noindent
(4) \textit{Filter} is a function:
\begin{align} \sigma : \Big((\mathbb{Z}^{*})^k  \rightarrow \{\texttt{true}, \texttt{false}\} \Big) \nonumber \rightarrow \left(R^{(k, r, \textbf{b})} \rightarrow R^{(k, r, \textbf{b})} \right) \nonumber
\end{align}
\noindent $\sigma_{\texttt{(boolFunc)}}$ returns a function that accepts a tensor relation and filters each of the tuples in the tensor relation by applying \texttt{boolFunc} to the keys in the tuples.

\noindent
(5) \textit{Transform} is a function:
\begin{align}\lambda : \Big( T^{(r_i, \textbf{b}_i)} \rightarrow T^{(r_o, \textbf{b}_o)} \Big) \rightarrow \left(R^{(k, r_i, \textbf{b}_i)} \rightarrow R^{(k, r_o, \textbf{b}_o)} \right) \nonumber 
\end{align}
\noindent $\lambda _{\texttt{(transformFunc)}}$ returns a function that accepts a tensor relation and applies the kernel function \texttt{transformFunc} to the array in each tuple from the tensor relation.

For an example of the rekey, filter and transform operations, assume we have a kernel operation \texttt{diag} that diagonalizes a matrix block, a function \texttt{isEq}$(\langle \texttt{k}_0, \texttt{k}_1) \rangle \mapsto \texttt{k}_0 = \texttt{k}_1$ that accepts a key and returns true if entries in position $0$ and $1$ in the key are identical to one another, and a function \texttt{getKey0}$(\langle \texttt{k}_0, \texttt{k}_1 \rangle) \mapsto \langle \texttt{k}_0 \rangle$ that returns the first dimension of a key.
We can use these functions along with filter, rekey, and transform to diagonalize a matrix \textbf{A} represented as a tensor relation $\texttt{R}_A$, by first examining the keys to remove all pairs that do not contain entries along the diagonal, and then diagonalizing the resulting arrays:
$$\lambda_{\left(\texttt{diag}\right)} \left( \rekey_{\left(\texttt{getKey0}\right)} \left( \sigma _{\left(\texttt{isEq}\right)} \left( \texttt{R}_A\right)\right)\right).$$

In addition, there are a number of operations that can be used to alter the organization of arrays within a tensor relation.  This allows the manipulation of how a tensor is represented as a tensor relation.  For this purpose, we have tile and concat:

\noindent
(6) \textit{Tile}:
\begin{align}
    \tile : \left(\mathbb{Z}^{*} \times \mathbb{Z}^{*} \right) \rightarrow \left( R^{(k, r, \textbf{b})}  \rightarrow R^{(k+1, r, \textbf{b}')}\right) \nonumber
\end{align}
$\tile _{\texttt{(tileDim, tileSize)}}$ returns a function that decomposes (or tiles) each array along a dimension \texttt{tileDim} to arrays of the target \texttt{tileSize} (by applying the \texttt{arrayTileOp} function on the array). As a result, a new key dimension is created, that effectively counts which tile the tuples holds along the tiling dimension.

For example, consider the matrix \textbf{B},
$$\textbf{B} = \begin{bmatrix}
1 & 2 & 5 & 6 & 9 & 10 & 13 & 14 \\
3 & 4 & 7 & 8 & 11 & 12 & 15 & 16\\
\end{bmatrix},$$
partitioned by columns and stored in tensor relation:
\begin{align}
\texttt{R}_B = \left\{ \left( \langle 0 \rangle, \begin{bmatrix} 1 & 2 & 5 & 6 \\ 3 & 4 & 7 & 8 \end{bmatrix}  \right),
        \left( \langle 1 \rangle, \begin{bmatrix} 9 & 10 & 13 & 14 \\ 11 & 12 & 15 & 16 \end{bmatrix}  \right)
\right\} 
\nonumber
\end{align}
If we make the call $\tile_{\left(1, 2\right)} \left( \texttt{R}_B\right)$, we will decompose each array along dimension $1$, creating one new array for each two columns.
In addition, a new key dimension is created, that effectively counts which tile the pair holds along the tiling dimension:
\begin{align}
    \tile_{\left(1, 2\right)} \left( \texttt{R}_B\right)  = 
    \Bigg\{ &\left( \langle 0, 0 \rangle, \begin{bmatrix} 1 & 2 \\ 3 & 4 \end{bmatrix}  \right),
        \left( \langle 0, 1 \rangle, \begin{bmatrix} 5 & 6 \\ 7 & 8 \end{bmatrix}  \right), \nonumber \\
        &\left( \langle 1, 0 \rangle, \begin{bmatrix} 9 & 10 \\ 11 & 12 \end{bmatrix}  \right),
        \left( \langle 1, 1 \rangle, \begin{bmatrix} 13 & 14 \\ 15 & 16 \end{bmatrix}  \right)
\Bigg\}. \nonumber \end{align}
\noindent
We may sometimes find ourselves in a situation where it is necessary to manipulate the key in each pair in a tensor relation so that the
key is consistent with the desired interpretation.  For example, the tensor relation $\texttt{R}_B$ defined above can represent a matrix with eight columns and
two rows, so $\tile_{\left(1, 2\right)} \left( \texttt{R}_B\right)$ is inconsistent with this, logically representing a matrix having four columns and four rows. For this purpose, we can leverage the $\rekey$ operator as we defined before. 

For example, we can rekey the output of $\tile_{\left(1, 2\right)} \left( \texttt{R}_B\right)$ so that logically, it corresponds to a two-by-eight matrix:
$$\rekey_{\left( \langle \texttt{k}_0, \texttt{k}_1\rangle \rightarrow \langle 2\texttt{k}_0+\texttt{k}_1 \rangle\right)} \left(\tile_{\left(1, 2\right)} \left( \texttt{R}_B\right) \right)$$
This will result in:
\begin{align}\Bigg\{ &\left( \langle 0 \rangle, \begin{bmatrix} 1 & 2 \\ 3 & 4 \end{bmatrix}  \right),
        \left( \langle 1 \rangle, \begin{bmatrix} 5 & 6 \\ 7 & 8 \end{bmatrix}  \right), 
        \left( \langle 2 \rangle, \begin{bmatrix} 9 & 10 \\ 11 & 12 \end{bmatrix}  \right),
        \left( \langle 3 \rangle, \begin{bmatrix} 13 & 14 \\ 15 & 16 \end{bmatrix}  \right)
\Bigg\} \nonumber \end{align}

Finally, we have the ability to undo a tiling.

\noindent
(7) \textit{Concat}:
\begin{align} \concat :  \left(\mathbb{Z}^{*} \times \mathbb{Z}^{*} \right) \rightarrow \left( R^{(k, r, \textbf{b})}  \rightarrow R^{(k-1, r, \textbf{b'})}\right)\nonumber \end{align}
\noindent $\concat_{\texttt{(keyDim, arrayDim)}}$ is an inverse to tile, which first groups all pairs in the relation using all of the key dimensions \emph{other than} \texttt{keyDim}, then concatenates all of the arrays in each group along \texttt{arrayDim}, with the concatenation ordering provided by \texttt{keyDim}. 

A call to $\concat_{\left (1, 1\right)} \left(\tile_{\left(1, 2\right)} \left( \texttt{R}_B\right)\right)$ first groups all pairs in $\tile _{\left(1, 2\right)} \left( \texttt{R}_B\right)$ using all of the key dimensions other than key dimension $1$, and then concatenates the arrays in each group along array dimension $1$, with the ordering provided by key dimension $1$.  Hence, this computation simply results in the recovery of $\texttt{R}_B.$

\subsection{Integrity Constraints and Closedness}
\label{sec:tra_discussion}

There are two important integrity constraints that each tensor relation must follow: uniqueness of keys, and a lack of ``holes'' in the tensor relation.  The primary reason for defining these constraints is facilitating easy, cost-based optimization.  With such constraints, cardinality estimation, one of the most vexing problems in relational optimization, goes away---see Section \ref{sec:cost}. Further, neither is particularly burdensome when expressing computations using the TRA.  In fact, if the interpretation of a tensor relation of type $R^{(k, r, \textbf{b})}$ is that it represents a $r$-dimensional tensor decomposed into chunks, these constraints are quite natural:

\begin{itemize}
\item \emph{Uniqueness}: every key should be unique in a tensor relation.
\item \emph{Continuity}: there are no ``holes''. Given a tensor relation $\texttt{R}$, of key-arity $k$, define the \emph{frontier} of $\texttt{R}$ as $\frontier(\texttt{R})$. $\textbf{f} = \frontier(\texttt{R})$ is a $k$-dimensional vector that bounds all keys in $\texttt{R}$. That is, for each key vector $\textbf{k}$ in $\texttt{R}$, $\textbf{k} < \textbf{f}$. 
Further, the frontier is the ``smallest'' vector bounding $\texttt{R}$, in that for any other vector $\textbf{f}'$ bounding $\texttt{R}$, $\textbf{f} \leq \textbf{f}'$.  \emph{Continuity} requires that for any vector $\textbf{k} < \textbf{f}$, some tuple in $\texttt{R}$ have the key $\textbf{k}$. 
\end{itemize}

It is easy to show that for the majority of TRA operations---the exceptions being the rekey and filter operations---tensor relations are closed.  That is, if the input(s) are tensor relation(s) that obey uniqueness and continuity, then the output must be a tensor relation that obeys these constraints.  Filtering a tensor relation or altering the keys can obviously violate the constraints, where the the former probably leads to holes in the resulting relation, and the latter can result in repeated key values. Analyzing a TRA expression to automatically detect whether it can violate these constraints is left as future work; we conjecture that if the filtering predicate (or re-keying computation) are limited to simple arithmetic expressions, it may be possible to check for closedness using an SMT solver \cite{de2008z3}.

\section{Implementation Algebra}

We now describe TRA's \textit{implementation algebra} (IA) that is suitable for execution in a parallel/distributed environment.

In IA, we extend each \texttt{(key, array)} tuple in a tensor relation with an additional \texttt{site} attribute, so that a \emph{physical tensor relation} $\phys{R}$ will consist of triples:
$$\texttt{(key, array, site)}.$$
The \texttt{site} attribute takes a value in $\{1,2,...,s\}$ where $s$ is the number of computation sites. Conceptually, the \texttt{site} value indicates the location where the tuple is stored; this could be a machine in a distributed cluster, or a compute unit like a GPU. 

Each physical tensor relation can map a particular $\texttt{(key, array)}$ pair to one or more sites. There are a few especially important mappings, recognized by the predicates $\textsc{All}()$ and $\textsc{Part}_D()$:

\begin{enumerate}
    \item If $\textsc{All}(\phys{R}) = \textrm{ true}$, it indicates that if we project away the \texttt{array} attribute, the resulting set will take the value: $$\{\textbf{k} \textrm{ s.t. } \textbf{k} \le \textsc{Front}(\phys{R}) \times \{1...s\}\}$$
    where $ \textsc{Front}(\phys{R})$ is the frontier of $\phys{R}$ (the frontier of a physical tensor relation is defined as in a ``regular'' tensor relation). In other words, this means that each possible  \texttt{(key, array)} tuple in $\phys{R}$ appears at all sites.
    
 \item If $\textsc{Part}_D(\phys{R}) = \textrm{ true}$ for some set $D \subseteq \{1...k\}$, it indicates that: (i) for a given \texttt{key} value, there is only one tuple in $\phys{R}$ and (ii) two tuples with the same key values for all dimensions in $D$ must be found on the same site.  In other words, $\phys{R}$ is partitioned according to the key dimensions in the set $D$. 
\end{enumerate}

We are ready to describe the IA. Let $\mathcal{R}^{(k, r, \textbf{b}, s)}$ specify the set of all valid physical tensor relations with key-arity of dimension $k$, storing arrays of type $T^{(r, \textbf{b})}$, and partitioned across $s$ sites.

The first two operations are concerned with manipulating the assigning of tuples in a physical relation to sites, while the later four operations operate over the \texttt{key} and \texttt{array} attributes.  

\noindent
(1) \textit{Broadcast} is defined as
\begin{align}
\broadcast : \mathcal{R}^{(k, r, \textbf{b}, s)} \rightarrow \mathcal{R}^{(k, r, \textbf{b}, s)} 
\nonumber
\end{align}

\noindent Given a physical tensor relation, $\broadcast$ simply ensures that each tuple takes each site value, so that (i) the set of \texttt{(key, array}) pairs is unchanged after $\broadcast$, but (ii) in any physical relation $\phys{R}$ output from a broadcast, $\textsc{All}(\phys{R}) = \textrm{ true}$.

\noindent
(2) \textit{Shuffle} is defined as:
\begin{align}
\shuffle : 2^{\{1...k\}} \rightarrow 
\left(\mathcal{R}^{(k, r, \textbf{b}, s)}  \rightarrow \mathcal{R}^{(k, r, \textbf{b}, s)}\right)\nonumber
\end{align}
\noindent $\shuffle_{(\texttt{partDims})}$ is a function that accepts a set of key dimensions, and returns a function that accepts physical tensor relation, and then repartitions the physical tensor relation, so that (i) the set of \texttt{(key, array}) pairs is unchanged after $\shuffle$, but (ii) in any physical relation $\phys{R}$ output from a shuffle, $\textsc{Part}_{\texttt{partDims}}(\phys{R}) = \textrm{ true}$.

\noindent
(3) \textit{Local join} is an extension of the TRA's join operation:
\begin{align} 
&\localjoin : \bigg( (\mathbb{Z}^{*})^{g} \times (\mathbb{Z}^{*})^{g} \times \left(T^{(r_l, \textbf{b}_l)} \times T^{(r_r, \textbf{b}_r)} \rightarrow T^{(r_o, \textbf{b}_o)}\right)\bigg) \nonumber \\
&\rightarrow \left(\mathcal{R}^{(k_l, r_l, \textbf{b}_l, s)} \times \mathcal{R}^{(k_r, r_r, \textbf{b}_r, s)} \rightarrow \mathcal{R}^{(k_l + k_r - g, r_o, \textbf{b}_o, s)} \right) \nonumber 
\end{align}
\noindent Similar to TRA join ($\Join$), $\localjoin_{\texttt{(joinKeysL, joinKeysR, projOp)}}$ takes as input a set of key dimensions to join on from the left and from the right, as well as a kernel operation to run over all \texttt{(leftArray, rightArray)} pairs that are created during the join. 
The key difference that the local join combines \emph{only} on pairs from the left and right inputs that have the \textbf{same} \texttt{site} values. If two tuples successfully join, the corresponding output tuple will have the \texttt{site} value as those input tuples.

\noindent
(4) \textit{Local aggregation} is an extension of TRA aggregation:
\begin{align}
\localagg : \left( (\mathbb{Z}^{*})^k \times \left(T^{(r, \textbf{b})} \times T^{(r, \textbf{b})}
\rightarrow T^{(r, \textbf{b})}\right)\right) \nonumber \\ \rightarrow \left(\mathcal{R}^{(k, r, \textbf{b}, s)}  \rightarrow \mathcal{R}^{(g, r, \textbf{b}, s)} \right) \nonumber \end{align}
\noindent Like TRA aggregation ($\Sigma$), $\localagg_{\texttt{(groupByKeys, aggOp)}}$ takes as input a list of key dimensions to aggregate over \texttt{groupByKeys} as well as a kernel function \texttt{aggOp}. However, it returns a function that takes as input a physical tensor relation, groups the arrays in the relation based upon the indicated key values \emph{and} the site value, and applies \texttt{aggOp} to the arrays in the group. Each output tuple in the resulting, physical tensor relation will take its \texttt{site} value from the \texttt{site} value of the set of input tuples that were aggregated to produce it.

\noindent
(5) IA has a filter:
\begin{align} 
\localfilter : \Big((\mathbb{Z}^{*})^g  \rightarrow \{\texttt{true}, \texttt{false}\} \Big) \rightarrow \left(\mathcal{R}^{(k, r, \textbf{b}, s)}  \rightarrow \mathcal{R}^{(k, r, \textbf{b}, s)} \right) \nonumber
\end{align}
\noindent The only difference is that each accepted input tuple's \texttt{site} value is carried through the filter.

\noindent
(6) Map provides two functionalities:
\begin{align}
\localmap : {\Big( \left((\mathbb{Z}^{*})^{k_i}  \rightarrow \big((\mathbb{Z}^{*})^{k_o} \big)^m \right) \times \left(T^{(r_i, \textbf{b}_i)} \rightarrow \big( T^{(r_o, \textbf{b}_o)}\big)^m\right) \Big) } \nonumber\\
\rightarrow \left(\mathcal{R}^{(k_i, r_i, \textbf{b}_i, s)} \rightarrow \mathcal{R}^{(k_o, r_o, \textbf{b}_o, s)} \right) \nonumber 
\end{align}

\noindent $\localmap_{\texttt{(keyMapFunc,arrayMapFunc)}}$ is a multi-map.  It returns a function that applies \texttt{keyMapFunc} to each \texttt{key} value in the input and applies \texttt{arrayMapFunc} to each \texttt{array} value in the input.  Both \texttt{keyMapFunc} and \texttt{arrayMapFunc} return $m$ output tuples per input tuple; the \texttt{site} value is simply copied from input to output. We subsequently call $m$ the \emph{arity} of \texttt{keyMapFunc}/\texttt{arrayMapFunc}.  In most cases the arity of these functions will be one, but on some cases (such as replication-based matrix multiply, see Section 4.2.2), the arity will be greater.

\section{Compilation and Optimization}
To distribute tensor-based computations so that they can run efficiently requires an optimization framework.  We consider three core questions related to actually distributing a computation specified in the TRA: (1) How is that TRA computation compiled into an equivalent statement in IA?  (2) What are a set of equivalence rules that allow computations in IA to be re-written, so as to produce different implementations that are known to produce the same results, but may run more efficiently? And (3) How to cost those different, equivalent implementations, so that a search strategy can be used to choose the most efficient one?

{
\begin{table*}[]
\small
\centering
\caption{Translation from TRA to IA.}
\begin{tabular}{|l|l|}
\hline
\textbf{TRA expression} & \textbf{Corresponding IA}\\\hline
$\Sigma _{\texttt{(groupByKeys, aggOp)}}\left(\texttt{R}\right)$ &
$\localagg_{\texttt{(groupByKeys, aggOp)}}\left(\shuffle_{\texttt{(groupByKeys)}}\left(\phys{R}\right)\right)$ \\\hline

$\Join_{\texttt{(joinKeysL, joinKeysR, projOp)}}\left(\texttt{R}_l, \texttt{R}_r\right)$ &
$\localjoin_{\texttt{(joinKeysL, joinKeysR, projOp)}}\left(\broadcast\left(\phys{R}_l\right),\phys{R}_r\right)$ \\\hline

$\rekey_{\texttt{(keyFunc)}}\left(\texttt{R}\right)$ &
$\localmap_{\texttt{(keyFunc,idOp)}}\left(\phys{R}\right)$\\\hline

$\sigma_{\texttt{(boolFunc)}}\left(\texttt{R}\right)$ & $\localfilter_{(\texttt{boolFunc})}\left(\phys{R}\right)$\\\hline

$\lambda _{\texttt{(transformFunc)}}\left(\texttt{R}\right)$ &  
$\localmap_{\texttt{(idOp,transformFunc)}}\left(\phys{R}\right)$\\\hline

$\tile _{\texttt{(tileDim, tileSize)}}\left(\texttt{R}\right)$ &
$\localmap_{\texttt{(keyTileOp(tileDim, tileSize), arrayTileOp(tileDim, tileSize))}}\left(\phys{R}\right)$ \\\hline

$\concat_{\texttt{(keyDim, arrayDim)}}\left(\texttt{R}\right)$ &
$\localagg_{(\langle \texttt{keyDim}\rangle^c, \texttt{arrayConcatOp})}\left(\shuffle_{\langle \texttt{keyDim}\rangle^c}\left(\phys{R}\right)\right)$ \\\hline
\end{tabular}
\label{tab:compile}
\end{table*}
}

\subsection{Compiling the TRA}

A complete set of rules mapping from TRA operations to IA operations are listed in Table \ref{tab:compile}\footnote{In Table \ref{tab:compile}, tensor relations \texttt{R}, $\texttt{R}_l$ and $\texttt{R}_r$ are stored as the corresponding physical tensor relations $\phys{R}$, $\phys{R}_l$ and $\phys{R}_r$; \texttt{idOp} represents an identity map for key or array; \texttt{arrayTileOp} is a kernel function splitting the array to chunks of the indicated size on the indicated dimension; $\texttt{arrayConcatOp}$ reverses this. \texttt{keyTileOp} is similar to \texttt{arrayTileOp}, but operates on keys; $\langle \texttt{keyDim}\rangle^c$ represents the complement set of $\langle \texttt{keyDim}\rangle$.}.
Note that though there can be multiple IA computations for a given TRA computation, the compiler will generate one of such IA computation as the initial plan, and an optimizer will typically be responsible for applying a search algorithm to produce the optimal physical plan represented by IA.

\subsection{Equivalence Rules}

Once a (possibly inefficient) computation in IA is produced, it can be optimized via the application of a set of equivalence rules.  An \emph{equivalence rule} for IA expressions holds if, for any input physical tensor relations, the two expressions produce equivalent outputs---two physical tensor relations are said to be equivalent if they contain the same set of \texttt{(key, array)} pairs after projecting away the \texttt{site} attribute.

There are two classes of equivalence rules we consider: \emph{simple equivalence rules} which are often extensions of classic relational equivalence rules (e.g., commutative property of selections), and \emph{domain-specific equivalence rules} that are more complex transformations that always hold, and tend to be useful for mathematical computations, such as matrix multiplications.

\subsubsection{Simple Equivalence Rules}

In Table \ref{tab:rules}, we give an extensive list of two types of simple equivalence rules: (i) those based on kernel function composition, and (ii) equivalence rules based on optimization of re-partitions. 

Kernel function composition targets the order or location of the application of kernel functions in order to reduce the computation load and memory consumption. This is closely related to the idea of ML operator fusion, which has been explored in systems such as TVM \cite{chen2018tvm} (though TVM does not consider the sort of distributed computations considered here). Re-partition rules formalize notions of distributed query optimization over tensor relations, and are primarily designed to reduce communication.

{
\begin{table*}[]
\small
\centering
\caption{Simple equivalence rules for kernel function composition and re-partition enumeration.}
\begin{tabular}{|l|}
\hline
\textbf{Kernel function composition based rules}:\\\hline

\textbf{R1-1.} Filter operations can be merged. \\
For a physical relation $\phys{R}$: \\
$\localfilter_{\left(\texttt{boolFunc1}\right)}\left(\localfilter_{\left(\texttt{boolFunc2}\right)} \left(\phys{R}\right)\right) \equiv \localfilter_{\left(\texttt{boolFunc1}\wedge\texttt{boolFunc2}\right)} \left(\phys{R}\right).$ \\\hline

\textbf{R1-2.} Map operations can be merged, if the output arity of the key and array mapping functions is one. \\
For a physical relation $\phys{R}$: \\
$\localmap_{\left(\texttt{keyMapFunc1, arrayMapFunc1}\right)}\left(\localmap_{\left(\texttt{keyMapFunc2, arrayMapFunc2}\right)} \left(\phys{R}\right)\right) \equiv \localmap_{\left(\texttt{keyMapFunc1}\circ\texttt{keyMapFunc2}, \texttt{arrayMapFunc1}\circ\texttt{arrayMapFunc2}\right)} \left(\phys{R}\right)$. \\\hline

\textbf{R1-3.} Map and filter are commutative if \texttt{keyMapFunc} is an identify function (\texttt{idOp}). \\
For a physical relation $\phys{R} \in \mathcal{R}^{(k, r, \textbf{b}, s)}$, if $\forall \ \texttt{key} \in \left(\mathbb{Z}^{*}\right)^k$, $\texttt{keyMapFunc}(\texttt{key})=\texttt{key}$: \\
$\localmap_{\left(\texttt{keyMapFunc, arrayMapFunc}\right)}\left(\localfilter_{\left(\texttt{boolFunc}\right)} \left(\phys{R}\right)\right) \equiv \localfilter_{\left(\texttt{boolFunc}\right)}\left(\localmap_{\left(\texttt{keyMapFunc, arrayMapFunc}\right)}\left(\phys{R}\right)\right)$. \\\hline

\textbf{R1-4.} The \texttt{arrayMapFunc} in map can be composed with local aggregation if \texttt{keyMapFunc} is an identify function (\texttt{idOp}): \\
For a physical relation $\phys{R} \in \mathcal{R}^{(k, r, \textbf{b}, s)}$, if $\forall \ \texttt{key} \in \left(\mathbb{Z}^{*}\right)^k$, $\texttt{keyMapFunc}(\texttt{key})=\texttt{key}$:\\
$\localmap_{\left(\texttt{keyMapFunc, arrayMapFunc}\right)}\left(\localagg_{\left(\texttt{groupByKeys, aggOp}\right)}\left(\phys{R}\right)\right) \equiv \localagg_{\left(\texttt{groupByKeys, arrayMapFunc} \circ \texttt{aggOp}\right)}\left(\phys{R}\right)$ \\
And if the kernel function \texttt{arrayMapFunc} and \texttt{aggOp} is distributive; that is, $\forall \ \texttt{array}_1,\texttt{array}_2 \in T^{(r, \textbf{b})}$, \\ $\texttt{arrayMapFunc} \left( \texttt{aggOp}\left(\texttt{array}_1,\texttt{array}_2\right)\right)=\texttt{aggOp} \left( \texttt{arrayMapFunc}\left(\texttt{array}_1\right),\texttt{arrayMapFunc}\left(\texttt{array}_2\right)\right)$: \\
$\localmap_{\left(\texttt{keyMapFunc, arrayMapFunc}\right)}\left(\localagg_{\left(\texttt{groupByKeys, aggOp}\right)}\left(\phys{R}\right)\right) \equiv \localagg_{\left(\texttt{groupByKeys, aggOp} \circ \texttt{arrayMapFunc}\right)}\left(\phys{R}\right)$ \\\hline

\textbf{R1-5.} The \texttt{boolFunc} in filter can be composed with local aggregation if the kernel function only depends on \texttt{groupByKeys}.\\
For a physical relation $\phys{R} \in \mathcal{R}^{(k, r, \textbf{b}, s)}$, $\forall \ \texttt{key}_1,\texttt{key}_2 \in \left(\mathbb{Z}^{*}\right)^k$, if \\
$\texttt{boolFunc}\left(\Pi_{\texttt{groupByKeys}}\left(\texttt{key}_1\right)\right)=\texttt{boolFunc}\left(\Pi_{\texttt{groupByKeys}}\left(\texttt{key}_2\right)\right)\Rightarrow \texttt{boolFunc}\left(\texttt{key}_1\right)=\texttt{boolFunc}\left(\texttt{key}_2\right)$:\\
$\localfilter_{\left(\texttt{boolFunc}\right)}\left(\localagg_{\left(\texttt{groupByKeys, aggOp}\right)}\left(\phys{R}\right)\right) \equiv \localagg_{\left(\texttt{boolFunc(groupByKeys), aggOp}\right)}\left(\phys{R}\right)$ \\\hline

\textbf{R1-6.} The kernel function in local filter can be pushed down with local join if the \texttt{boolFunc} only checks the joined keys . \\
For physical relations $\phys{R}_l$ and $\phys{R}_r$:\\
$\localfilter_{\left(\texttt{boolFunc}\right)}\left(\localjoin_{\left(\texttt{joinKeysL, joinKeysR, projOp}\right)}\left(\phys{R}_l,\phys{R}_r\right)\right) \equiv \localjoin_{\left(\texttt{joinKeysL, joinKeys, projOp}\right)}\left(\localfilter_{\left(\texttt{boolFunc}\right)}\left(\phys{R}_l\right),\localfilter_{\left(\texttt{boolFunc}\right)}\left(\phys{R}_r\right)\right)$.\\

\hline

\textbf{R1-7.} The kernel function in local map can be composed with local join, if the output arity of the key and array mapping functions is one.\\
For physical relations $\phys{R}_l$ and $\phys{R}_r$:\\
$\localmap_{\left(\texttt{keyMapFunc, arrayMapFunc}\right)}\left(\localjoin_{\left(\texttt{joinKeysL, joinKeysR, projOp}\right)}\left(\phys{R}_l,\phys{R}_r\right)\right) \equiv \localjoin_{\left(\texttt{keyMapFunc(joinKeysL), keyMapFunc(joinKeysR), arrayMapFunc} \circ \texttt{projOp}\right)}\left(\phys{R}_l,\phys{R}_r\right)$.\\
And if the kernel function \texttt{arrayMapFunc} and \texttt{projOp} is distributive, that is, $\forall \ \texttt{array}_1,\texttt{array}_2 \in T^{(r, \textbf{b})}$, \\ $\texttt{arrayMapFunc}: \left( \texttt{projOp}\left(\texttt{array}_1,\texttt{array}_2\right)\right)=\texttt{projOp} \left( \texttt{arrayMapFunc}\left(\texttt{array}_1\right),\texttt{arrayMapFunc}\left(\texttt{array}_2\right)\right)$: \\
$\localmap_{\left(\texttt{keyMapFunc, arrayMapFunc}\right)}\left(\localjoin_{\left(\texttt{joinKeysL, joinKeysR, projOp}\right)}\left(\phys{R}_l,\phys{R}_r\right)\right) \equiv \localjoin_{\left(\texttt{keyMapFunc(joinKeysL), keyMapFunc(joinKeysR), projOp} \circ \texttt{arrayMapFunc}\right)}\left(\phys{R}_l,\phys{R}_r\right)$.\\

\hline
\textbf{Re-partition based rules}: \\\hline

\textbf{R2-1.} Only the final broadcast/shuffle in a sequence of broadcast/shuffle operations is needed.\\
For a physical relation $\phys{R}$: \\
$\broadcast\left(\broadcast\left(...\broadcast\left(\phys{R}\right)\right)\right) \equiv \broadcast\left(\phys{R}\right)$\\
$\shuffle_{\left(\texttt{partDims}_n\right)}\left(...\shuffle_{\left(\texttt{partDims}_2\right)}\left(\shuffle_{\left(\texttt{partDims}_1\right)}\left(\phys{R}\right)\right)\right) \equiv \shuffle_{\left(\texttt{partDims}_n\right)}\left(\phys{R}\right)$.\\\hline

\textbf{R2-2.} The re-partition operations are commutative with the local filter operation.\\
For a physical relation $\phys{R}$: \\
$\broadcast\left(\localfilter_{\left(\texttt{boolFunc}\right)}\left(\phys{R}\right) \right) \equiv \localfilter_{\left(\texttt{boolFunc}\right)}\left(\broadcast\left(\phys{R}\right)\right)$; \\
$\shuffle_{\left(\texttt{partDim}\right)}\left(\localfilter_{\left(\texttt{boolFunc}\right)}\left(\phys{R}\right) \right) \equiv \localfilter_{\left(\texttt{boolFunc}\right)}\left(\shuffle_{\left(\texttt{partDims}\right)}\left(\phys{R}\right)\right)$. \\\hline

\textbf{R2-3.} The re-partition operations are commutative with the local map operation. \\
For a physical relation $\phys{R}$: \\
$\broadcast\left(\localmap_{\left(\texttt{keyMapFunc,arrayMapFunc}\right)}\left(\phys{R}\right) \right) \equiv \localmap_{\left(\texttt{keyMapFunc,arrayMapFunc}\right)}\left(\broadcast\left(\phys{R}\right)\right)$;\\
And, if \texttt{keyMapFunc} is the identity function:\\
$\shuffle_{\left(\texttt{partDims}\right)}\left(\localmap_{\left(\texttt{keyMapFunc,arrayMapFunc}\right)}\left(\phys{R}\right) \right) \equiv \localmap_{\left(\texttt{keyMapFunc,arrayMapFunc}\right)}\left(\shuffle_{\left(\texttt{partDims}\right)}\left(\phys{R}\right)\right)$.\\\hline

\textbf{R2-4.} A shuffle can be avoided if the physical relation is already partitioned by a local aggregation's \texttt{groupByKeys}. \\
For a physical relation $\phys{R}$, if $\texttt{partDims} \subseteq \texttt{groupByKeys}$: \\
$\localagg_{\left(\texttt{groupByKeys, aggOp}\right)}\left( \shuffle_{\left(\texttt{partDims}\right)}\left(\phys{R}\right)\right) \equiv \localagg_{\left(\texttt{groupByKeys, aggOp}\right)}\left(\phys{R}\right)$ \\ \hline

\textbf{R2-5.} An aggregation can be split to two phases, if the physical relation is only partially partitioned. \\
For a physical relation $\phys{R}$, if $\texttt{groupByKeys} \subset \texttt{partDims}$:\\
$\localagg_{\left(\texttt{groupByKeys, aggOp}\right)} \left( \shuffle _{\left(\texttt{partDims}\right)}\left(\phys{R}\right)\right) \equiv \localagg_{\left(\texttt{groupByKeys, aggOp}\right)}\left(\shuffle_{\left(\texttt{partDims}\right)}\left(\localagg_{\left(\texttt{groupByKeys, aggOp}\right)} \left(\phys{R}\right)\right)\right)$. \\ \hline

\textbf{R2-6.} A Join $\Join$ defined by the TRA can be implemented in the following equivalent ways.\\
For physical relations $\phys{R}_l$ and $\phys{R}_r$:\\
$\localjoin_{\left(\texttt{joinKeysL, joinKeysR, projOp}\right)}\left(\broadcast\left(\phys{R}_l \right),\phys{R}_r\right)$
$\equiv \localjoin_{\left(\texttt{joinKeysL, joinKeysR, projOp}\right)}\left(\phys{R}_l,\broadcast\left(\phys{R}_r\right)\right)$ \\
$\equiv \localjoin_{\left(\texttt{joinKeysL, joinKeysR, projOp}\right)}\left(\shuffle_{\left(\texttt{joinKeysL}\right)}\left(\phys{R}_l \right),\shuffle_{\left(\texttt{joinKeysR}\right)}\left(\phys{R}_r\right)\right)$.\\
\hline

\textbf{R2-7.} The local join can be pushed through shuffle.\\
For physical relations $\phys{R}_l \in \mathcal{R}^{(k_l, r_l, \textbf{b}_l, s)}$ and $\phys{R}_r \in \mathcal{R}^{(k_r, r_r, \textbf{b}_r, s)}$, if $\texttt{partDims} \subseteq \texttt{joinKeysL}$:\\
$\shuffle_{\left(\texttt{partDims}\right)}\left(\localjoin_{\left(\texttt{joinKeysL, joinKeysR, projOp}\right)}\left(\shuffle_{\left(\texttt{joinKeysL}\right)}\left(\phys{R}_l\right), \shuffle_{\left(\texttt{joinKeysR}\right)}\left(\phys{R}_r\right) \right) \right) \equiv$ \\ $\localjoin_{\left(\texttt{joinKeysL, joinKeysR, projOp}\right)}\left(\shuffle_{\left(\texttt{joinKeysL}\right)}\left(\phys{R}_l\right), \shuffle_{\left(\texttt{joinKeysR}\right)}\left(\phys{R}_r\right)\right)$\\\hline 

\end{tabular}
\label{tab:rules}
\end{table*}
}

Such rules are surprisingly effective for optimizing distributed tensor manipulations.
Consider the example of extracting the diagonal elements of matrix $\textbf{X}$ plus matrix $\textbf{Y}$: $\texttt{diag}(\textbf{X}+\textbf{Y})$, where matrix $\textbf{X}$ and $\textbf{Y}$ are stored in physical tensor relations $\phys{R}_X$ and $\phys{R}_Y$. 
This computation can be represented by the following TRA expression, where $\texttt{isEq}(\langle \texttt{k}_0, \texttt{k}_1 \rangle) \mapsto \texttt{k}_0 = \texttt{k}_1$, $\texttt{merge}(\langle \texttt{k}_0, \texttt{k}_1 \rangle) \mapsto \langle \texttt{k}_0 \rangle$, \texttt{matAdd} is an element-wise sum of two arrays, and \texttt{diag} diagonalizes the array.  Then diag$(\textbf{X}+\textbf{Y})$ can be written as:

\begin{footnotesize}
$$\lambda_{\left(\texttt{diag}\right)} \left(\rekey_{(\texttt{merge})}\left( \sigma_{\left(\texttt{isEq}\right)} \left(\Join _{\left(\langle 0,1 \rangle, \langle 0,1 \rangle, \texttt{matAdd}\right)} \left(\texttt{R}_{X}, \texttt{R}_{Y} \right) \right)\right)\right).$$
\end{footnotesize}

This TRA expression can be translated to the IA expression: 

\begin{footnotesize}
$$\localmap_{\left(\texttt{idOp,diag}\right)}\left(\localmap_{\left(\texttt{merge,idOp}\right)}\left(\localfilter _{\left(\texttt{isEq}\right)} \left(\localjoin_{\left(\langle 0,1 \rangle, \langle 0,1 \rangle, \texttt{matAdd}\right)}\left(\broadcast\left(\phys{R}_{X}\right), \phys{R}_{Y}\right)\right)\right)\right).$$
\end{footnotesize}

We can apply the following equivalence rules for the above IA expression:
\begin{footnotesize}
\begin{align*}
&\localmap_{\left(\texttt{idOp,diag}\right)}\left(\localmap_{\left(\texttt{merge,idOp}\right)}\left(\localfilter _{\left(\texttt{isEq}\right)} \left(\localjoin_{\left(\langle 0,1 \rangle, \langle 0,1 \rangle, \texttt{matAdd}\right)}\left(\broadcast\left(\phys{R}_{X}\right), \phys{R}_{Y}\right)\right)\right)\right)\\
&\overset{R1-2}{\equiv}\: \localmap_{\left(\texttt{merge,diag}\right)}\left(\localfilter _{\left(\texttt{isEq}\right)} \left(\localjoin_{\left(\langle 0,1 \rangle, \langle 0,1 \rangle, \texttt{matAdd}\right)}\left(\broadcast\left(\phys{R}_{X}\right), \phys{R}_{Y}\right)\right)\right) \\
&\overset{R1-6}{\equiv}\:  \localmap_{\left(\texttt{merge,diag}\right)}\left(\localjoin_{\left(\langle 0,1 \rangle, \langle 0,1 \rangle, \texttt{matAdd}\right)}\left(\localfilter _{\left(\texttt{isEq}\right)} \left(\broadcast\left(\phys{R}_{X}\right)\right), \localfilter _{\left(\texttt{isEq}\right)} \left(\phys{R}_{Y}\right)\right)\right) \\
&\overset{R2-2}{\equiv}\:  \localmap_{\left(\texttt{merge,diag}\right)}\left(\localjoin_{\left(\langle 0,1 \rangle, \langle 0,1 \rangle, \texttt{matAdd}\right)}\left(\broadcast\left(\localfilter _{\left(\texttt{isEq}\right)} \left(\phys{R}_{X}\right)\right), \localfilter _{\left(\texttt{isEq}\right)} \left(\phys{R}_{Y}\right)\right)\right) \\
&\overset{R1-7}{\equiv}\:  \localjoin_{\left(\texttt{merge}\left(\langle0,1\rangle\right),\texttt{merge}\left(\langle0,1\rangle\right), \texttt{matAdd}\circ\texttt{diag} \right)}\left(\broadcast\left(\localfilter _{\left(\texttt{isEq}\right)} \left(\phys{R}_{X}\right)\right), \localfilter _{\left(\texttt{isEq}\right)} \left(\phys{R}_{Y}\right)\right).
\end{align*}
\end{footnotesize}

The transformation will significantly reduce both the communication overhead and the computation load: by applying R1-6, the \texttt{isEq} functions will be pushed down, this transformation not only reduces the input tuple pairs for the join to execute the \texttt{matAdd} function but also the enables reduction of communication overhead where the the filter operation is commuted with the broadcast operation by R2-2; lastly, R1-7 leverages the property that kernel functions \texttt{diag} and \texttt{matAdd} are distributive, as a result, addition will only be applied for the diagonal elements for the paired blocks after kernel function composition. 

\subsubsection{Domain-Specific Equivalence Rules} \label{sec:dom_rule}
Such rules encode specific knowledge from parallel and distributed computing algorithms.  Adding such rules to a system allows IA to have at its disposal common implementation strategies, that it can choose from in a cost-based manner.

We do not attempt to produce an exhaustive list of such rules, but rather we consider in detailing one example: distributed matrix multiplication over tensor relations $\texttt{R}_X$ and $\texttt{R}_Y$:

\begin{footnotesize}
$$\Sigma _{\left(\langle 0, 2 \rangle, \texttt{matAdd}\right)} \left(\Join _{\left(\langle 1 \rangle, \langle 0 \rangle, \texttt{matMul}\right)} \left(\texttt{R}_X, \texttt{R}_Y \right) \right).$$
\end{footnotesize}%

For physical tensor relations  $\phys{R}_X$ and $\phys{R}_Y$, using the rules of Table 1, this would be compiled into:

\begin{footnotesize}
$$\localagg_{\left(\langle 0, 2 \rangle, \texttt{matAdd}\right)}\left(\shuffle_{\left(\langle0,2\rangle\right)}\left( \localjoin_{\left(\langle1\rangle,\langle0\rangle,\texttt{matMul}\right)}\left(\broadcast\left(\phys{R}_X\right), \phys{R}_Y\right)\right)\right).$$ 
\end{footnotesize}%

This is a simple, broadcast-based matrix multiply.  Applying simple equivalence rules brings us to cross product-based matrix multiplication, which partitions $\phys{R}_X$ on columns, and $\phys{R}_Y$ on rows. The IA program is:

\begin{footnotesize}
\begin{align*}
   \localagg_{\left(\langle 0, 2 \rangle, \texttt{matAdd}\right)}\left(\shuffle_{\left(\langle0,2\rangle\right)}\left( \localjoin_{\left(\langle1\rangle,\langle0\rangle,\texttt{matMul}\right)}\left(\shuffle_{\left(\langle1\rangle\right)}\left(\phys{R}_X\right), 
   \shuffle_{\left(\langle0\rangle\right)}\left(\phys{R}_Y\right)\right)\right)\right).
\end{align*}
\end{footnotesize}%

However, more complicated schemes are possible, which are expressible in IA, but not derivable using the simple equivalence rules. For example, replication-based matrix multiplication can be viewed as a relational version of the 3D parallel matrix multiplication \cite{agarwal1995three}. The algorithm first replicates matrix \textbf{X} and \textbf{Y}’s blocks multiple times, viewing the result as a 3-D array, and shuffles them using the index of the corresponding voxel as a key; then each site joins the tuples with the same keys and performs local multiplications, aggregating to obtain the final results. 
If \texttt{xDups} is defined as $\frontier(\phys{R}_Y)[1]$ and \texttt{yDups} is $\frontier(\phys{R}_X)[0]$, the shuffle stage can be implemented in IA as:

\begin{footnotesize}
\begin{align*}
    \phys{R}_X^{*} = \shuffle_{\left(\langle0,2\rangle\right)}\left(\localmap_{\left(\texttt{insertDim}(2,\texttt{xDups}),\texttt{duplicate}(\texttt{xDups})\right)}\left(\phys{R}_X\right)\right)\\
    \phys{R}_Y^{*} = \shuffle_{\left(\langle0,2\rangle\right)}\left(\localmap_{\left(\texttt{insertDim}(0,\texttt{yDups}),\texttt{duplicate}(\texttt{yDups})\right)}\left(\phys{R}_Y\right)\right)
\end{align*}
\end{footnotesize}%

\noindent where kernel functions \texttt{insertDim} and \texttt{duplicate} add a new dimension, and duplicate each existing array the specified number of times. 
For example, applying $\localmap_{\left(\texttt{insertDim}(2,\texttt{xDups}),\texttt{duplicate}(\texttt{xDups})\right)}$ to the tensor relation 
\begin{align}\Bigg\{ &\left( \langle 0,0 \rangle, \begin{bmatrix} 1 & 2 \\ 3 & 4 \end{bmatrix}  \right),
        \left( \langle 0,1 \rangle, \begin{bmatrix} 5 & 6 \\ 7 & 8 \end{bmatrix}  \right),  \nonumber \\
        &\left( \langle 1,0 \rangle, \begin{bmatrix} 9 & 10 \\ 11 & 12 \end{bmatrix}  \right),
        \left( \langle 1,1 \rangle, \begin{bmatrix} 13 & 14 \\ 15 & 16 \end{bmatrix}  \right)
\Bigg\} \nonumber \end{align}
will produce:
\begin{align}\Bigg\{ &\left( \langle 0,0,0 \rangle, \begin{bmatrix} 1 & 2 \\ 3 & 4 \end{bmatrix}  \right),
        \left( \langle 0,0,1 \rangle, \begin{bmatrix} 1 & 2 \\ 3 & 4 \end{bmatrix}  \right), \nonumber \\
        &\left( \langle 0,1,0 \rangle, \begin{bmatrix} 5 & 6 \\ 7 & 8 \end{bmatrix}  \right),
        \left( \langle 0,1,1 \rangle, \begin{bmatrix} 5 & 6 \\ 7 & 8 \end{bmatrix}  \right), \nonumber \\
        &\left( \langle 1,0,0 \rangle, \begin{bmatrix} 9 & 10 \\ 11 & 12 \end{bmatrix}  \right),
        \left( \langle 1,0,1 \rangle, \begin{bmatrix} 9 & 10 \\ 11 & 12 \end{bmatrix}  \right), \nonumber \\
        &\left( \langle 1,1,0 \rangle, \begin{bmatrix} 13 & 14 \\ 15 & 16 \end{bmatrix}  \right),
        \left( \langle 1,1,1 \rangle, \begin{bmatrix} 13 & 14 \\ 15 & 16 \end{bmatrix}  \right)
\Bigg\} \nonumber \end{align}
\noindent Next we execute:

\begin{footnotesize}
$$\localagg_{\left(\langle 0, 2 \rangle, \texttt{matAdd}\right)}\left(\localjoin_{\left(\langle0,1,2\rangle,\langle0,1,2\rangle,\texttt{matMul}\right)}\left( \phys{R}_X^{*}, \phys{R}_Y^{*} \right)\right).$$
\end{footnotesize}%

The equivalence of these three implementations is an example of a set of domain-specific equivalence rules.

\subsection{Cost Model}
\label{sec:cost}

One of the beneficial aspects of the TRA is that cost-based optimization is much easier than that for classical relational algebra:  If the uniqueness and continuity constraints hold, tuple counts need not be estimated and can be computed exactly.

In the simple cost model presented here, we use the number of floating point values that must be transferred between sites as the cost metric. 
There are two reasons for this decision. 

Fist, the number of floating point operations in distributed ML computation is fixed.  For example, all of the classical distributed matrix multiply algorithms---2.5D \cite{solomonik2011communication}, SUMMA \cite{van1997summa}, etc., have the same floating point cost. While this is not a hard and fast rule---it is possible to push the filtering of tuples in a tensor relation past the application of a kernel function, which would change the number of floating-point operations---in many applications, network transfer is the dominant cost, and is a reasonable cost metric. 

Second, skew, which could slow down a computation with low network cost, is generally not an issue in a TRA computation, unlike in classical relational database system. The TRA continuity and uniqueness constraints imply that joins and aggregations cannot encounter skew. Consider a join. For two tuples $t_1$ and $t_2$ in tensor relation \texttt{R}, when that relation is joined with another relation \texttt{S}, the number of tuples from \texttt{S} that join with $t_1$ and $t_2$ must be the same. This, along with the fact that the TRA requires all arrays in a tensor relation to be of the same type, implies that skew will be very rare. In a TRA implementation, the only source of delay where the entire computation is blocked on a machine is likely to be a machine that is, simply stated, slower to perform computations than the others in the cluster. Such slowness may be due to hardware heterogeneity---a challenging issue for future work---or for unpredictable reasons, such as other workloads running on the machine, which are eventualities that cannot be planned for and must be handled by the runtime.

To compute the network transfer cost for a plan in IA, we need to be able to compute the frontier of each physical relation $\phys{R}$: $\textbf{f} = \textsc{Front}(\phys{R})$. The reason is that, assuming that uniqueness and continuity constraints hold, we can compute the number of floating point numbers in $\phys{R}$ using $\textbf{f}$. If $\phys{R}$ is of type $\mathcal{R}^{(k, r, \textbf{b}, s)}$, and $\textbf{f} = \textsc{Front}(\phys{R})$, then the number of tuples in the corresponding tensor relation is $n = \prod_i \textbf{f}_i$, and the number of floating point numbers in the tensor relation is $n \times \prod_i \textbf{b}_i$.  

Once the frontier is known, it is used to compute the transfer cost for each $\broadcast$ and $\shuffle$ operation.  The cost to broadcast a tensor relation of type $\mathcal{R}^{(k, r, \textbf{b}, s)}$ and having $f$ floating point numbers, is simply $f \times s$.  The cost to shuffle a tensor relation of $f$ floating point numbers is simply $f$. 


Thus, the task of costing a physical TRA plan reduces to computing the type of each intermediate relation, as well as its frontier.  

Computing the type is relatively easy: we work up from the leaves to the end result(s) of a physical plan, using the type signature of each of the physical operations (Section 4) to infer the type of each intermediate physical relation.

Computing the frontier in this way, working up from leaves to outputs, is also possible, but it requires a bit more thought.  We now consider how the frontier of an output is computed for each of the various operations in the physical algebra:

\begin{enumerate}
    \item $\localjoin_{\texttt{(joinKeysL, joinKeysR, projOp)}} \left(\phys{R}_l, \phys{R}_r\right)$. For local join, assuming that $\phys{R}_l$ and $\phys{R}_r$ have an appropriate partitioning to sites, let $\textbf{f}_l$ and $\textbf{f}_r$ be the left and right input frontiers of dimensionality $k_l$ and $k_r$, respectively.  Then the output frontier $\textbf{f}$ is computed as follows.  For $k < k_l$ and $k$ not in \texttt{joinKeysL}, $\textbf{f}[k]$ $=$ $\textbf{f}_l[k]$, as the frontier value for that dimension is inherited from the left.  For $k < k_l$ and where $k =$ \texttt{joinKeysL}$[i]$, $\textbf{f}[k]$ $=$ min$(\textbf{f}_l[k], \textbf{f}_r[i])$, as the frontier value for that dimension results from the join of the two relations.  And finally, for all other $k$, $\textbf{f}[k]$ is inherited from the corresponding dimension in the right frontier.
    
    \item $\localagg_{\texttt{(groupByKeys, aggOp)}} (\phys{R})$. For local aggregation, assuming an appropriate partitioning, let $\textbf{f}_i$ denote the input frontier, and let $n$ be the number of key dimensions in \texttt{groupByKeys}.  In this case, for $k \le n$, $\textbf{f}[k]$ $=$ $\textbf{f}_l[\texttt{groupByKeys}[k]]$.
    
    \item $\localfilter_{\texttt{(boolFunc)}}(\phys{R})$.  This performs a filter in the physical tensor relation $\phys{R}$.  For an $n$-dimensional input frontier $\textbf{f}_i$, for any $k \le n$, by definition:
    $$\qquad \quad \textbf{f}[k] = 1 + \textrm{max}\left\{\textbf{k}[k] \textrm{ s. t. } \textbf{k} < \textbf{f}_i  \textrm{ and } \texttt{boolFunc}(\textbf{k}) = \textrm{ true}\right\}.$$  
    That is, the $k$-th dimension in the frontier is inherited from the largest key value in that dimension accepted by \texttt{boolFunc}.  In many cases, especially if \texttt{boolFunc} consists of simple arithmetic expressions any comparisons, symbolic methods can be used to compute this.  But in practice, it may simply be easier to use a brute-force approach, where each key value is fed into \texttt{boolFunc} to compute the required maximum.  Since the size of a tensor relation is typically small---tens of thousands of tuples would be very large---this is a very practical approach.
    
    \item $\localmap_{\texttt{(keyMapFunc,arrayMapFunc)}}(\phys{R})$. Similarly, for an $n$-dimensional input frontier $\textbf{f}_i$, for any $k \le n$, by definition:
    $$\qquad \quad \textbf{f}[k]=1 + \textrm{max}\left\{\texttt{keyMapFunc}(\textbf{k})[k] \textrm{ s. t. } \textbf{k} < \textbf{f}_i\right\}.$$  
    Again, a brute force-approach is appropriate for computing the frontier in this case.
\end{enumerate}

\section{Evaluation}
{
\begin{table*}[t]
\small
\begin{center}
\caption{Distributed matrix multiply runtimes.}
\begin{tabular}{|c||c|c|c||c|c|c||c|c|c|}
\hline
 & \multicolumn{3}{|c||}{\texttt{Two General Matrices}} & \multicolumn{3}{|c||}{\texttt{A Common Large Dim}} & \multicolumn{3}{|c|}{\texttt{ Two Large Dims}} \\
 \hline
Cluster Size & 5 & 10 & 15 & 5 & 10 & 15 & 5 & 10 & 15 \\
\hline
BMM      &  61.18s  &  46.48s &  38.54s & 106.24s & 104.67s & 101.63s  &  57.23s &  37.60s &  31.64s  \\
CMM      &  63.14s  &  40.08s &  29.38s & 51.52s  &  30.58s &  23.09s  & 106.82s &  82.72s &  75.63s  \\
RMM      &  60.71s  &  43.56s &  44.55s & 91.19s  &  74.40s &  68.43s  &  59.91s &  41.12s &  33.26s \\
ScaLAPCK &  66.11s  &  37.05s &  28.30s &  83.96s &  58.17s &  35.45s  &  53.06s &  28.13s &  22.34s  \\
2.5D &  62.93s  &   29.60s  &  23.11s &  83.59s &   46.43s &  34.82s &  61.13s &   28.36s &  21.21s \\
Dask &  200.64s &   161.23s &  104.12s&  Fail   &   Fail   &  Fail &  Fail   &   Fail   &  Fail   \\
PETSc  &  1034.40s  &   535.85s  &  430.80s &  1071.62s  &   801.26s  &  550.74s &   1051.71s  &   810.61s  &  598.24s\\
\hline
\end{tabular}
\label{table:mm}
\end{center}
\end{table*}
}

{
\begin{table}[t]
\small
\begin{center}
\caption{Predicted costs for a 10-node cluster.}
\begin{tabular}{|c||c|c|c|}
\hline
                      & BMM & CMM & RMM \\
\hline
\texttt{Two General}           & $1.6\times10^{10}$    &  $1.6\times10^{10}$   &   $1.6\times10^{10}$  \\ 
\texttt{A Common Large Dim}    &  $6.4\times10^{10}$   &  $1.0\times10^{9}$    &   $6.4\times10^{10}$  \\
\texttt{Two Large Dims}       &  $8.0\times10^{9}$    &  $6.4\times10^{10}$   &   $8.0\times10^{9}$     \\
\hline
\end{tabular}
\label{table:mm_cost}
\end{center}
\end{table}
}

{
\begin{table*}[]
\small
\begin{center}
\caption{Nearest neighbor search runtimes.}
\begin{tabular}{|c||c|c|c|c|c|c||c|c|c|c|c|c|}
\hline
& \multicolumn{3}{|c|}{ \texttt{Wide  ($N=1.5\times10^5$)}} 
& \multicolumn{3}{|c||}{\texttt{Wide  ($N=1.5\times10^6$)}}& \multicolumn{3}{|c|}{ \texttt{Large ($D=3\times10^4$)}} & \multicolumn{3}{|c|}{\texttt{Large ($D=10^5$)}}\\ 
\hline
Cluster Size & 4 & 8 & 12 & 4 & 8 & 12 & 4 & 8 & 12 & 4 & 8 & 12\\
\hline
\texttt{Opt4Horizontal} &  5.64s  &   4.71s  &   3.36s  &  55.62s &  39.24s &  24.63s &  13.26s &  17.71s &  28.25s & 159.69s & 229.40s & 315.17s\\
\texttt{Opt4Vertical}   &  10.44s &   9.26s  &    9.88s & 120.09s & 112.69s & 108.59s &   5.93s &   4.52s &   3.82s &  57.81s &  35.61s &  26.43s\\
Single big machine &  5.31s  &   4.65s  &   3.33s &  51.54s &  35.82s &  23.01s &   5.21s &   4.41s  &   3.79s &  48.22s &  31.12s &  24.88s  \\
 Dask       &  485.87s   &   289.50s   &  223.55s  &  Fail   &   Fail   &  Fail &  437.31s   &   420.25s   &  381.10s &  Fail   &   Fail   &  Fail\\
\hline
\end{tabular}
\label{table:nns}
\end{center}
\end{table*}
}

{
\begin{table}[]
\small
\begin{center}
\caption{Predicted nearest neighbor search costs, 8 machines.}
\begin{tabular}{|c||c|c|}
\hline
                        &  \texttt{Opt4Horizontal} &  \texttt{Opt4Vertical}\\ \hline
\texttt{Wide data set}  &  $2.9 \times 10^8$       & $8.0 \times 10^{10}$ \\
\texttt{Large data set} &  $7.2 \times 10^{10}$    & $4.8\times10^9$  \\
\hline
\end{tabular}
\label{table:prednn}
\end{center}
\end{table}
}

The goal of our paper is to design a computational abstraction that could be exported by the back-end of a machine learning system. To be specific, we have detailed that: (1) such an abstraction should be expressive enough to express a variety of computations; (2) the computations expressed using the abstraction should be competitive with hand-coded or special-purpose solutions; and (3) the abstraction should be amenable to automatic optimization.

To determine whether the TRA meets these goals, we have implemented a simple Python back-end that exports the TRA/IA interface. Across three different large-scale ML computations, we experimentally evaluate this TRA-based back-end. To see whether a TRA-based back-end can provide suitable performance, we compare  with a number of other options, including hand-coded MPI solutions, high-performance libraries such as ScaLAPACK, distributed data science tools such as Dask, and ML systems such as TensorFlow and PyTorch.    To see whether it is amenable to automatic optimization, for each ML computation, we apply a series of transformations to obtain multiple implementations in the IA, and evaluate whether the cost model is able to predict which is preferable.

\noindent
\textbf{Benchmark Tasks.} (i) distributed matrix multiplication, (ii) distributed nearest neighbor search in a Riemannian metric space, and (iii) distributed stochastic gradient decent (SGD) in a two-layer, feed-forward neural network (FFNN).

\noindent
\noindent \textbf{TRA Implementation.} We implement an execution engine for the IA in Python. While it may seem surprising that Python is appropriate for implementing a relational engine, for even very large ML problems, the number of tuples in a TRA computation is small; most data are stored in the large arrays. Our Python execution engine makes heavy use of PyTorch to handle those arrays. PyTorch is used to actually execute the compute kernels on the various sites in a compute cluster, and our IA implementation uses PyTorch's optimized communication library to move the arrays stored in tensor relations between machines.  

\subsection{Matrix Multiplication}
Multiplication of $\textbf{A} \in \mathbb{R}^{I \times K }$ and $\textbf{B} \in \mathbb{R}^{K \times J}$ can be formalized:
\begin{footnotesize}
\begin{align*}
\Sigma _{\left(\langle 0, 2 \rangle, \texttt{matAdd}\right)} \left(\Join _{\left(\langle 1 \rangle, \langle 0 \rangle, \texttt{matMul}\right)} \left(  \texttt{R}_\textbf{A}, \texttt{R}_\textbf{B} \right) \right)
\end{align*}
\end{footnotesize}
\noindent where matrix $\textbf{A}$ and $\textbf{B}$ are stored in tensor relations $\texttt{R}_{\textbf{A}}$ and $\texttt{R}_{\textbf{B}}$.

To test the effectiveness of IA optimization, as others have done \cite{gu2017improving, han2019distme}, we consider three different multiplications: (i) general matrices ($I=K=J=4\times10^4$), (ii) matrices with a common large dimension ($K=6.4\times10^5$,  $I=J=10^4$), and (iii) matrices with two large dimensions ($I=J=8\times10^4$, $K=10^4$). Matrices are filled with random data following uniform distribution $\mathcal{U}(-1, 1)$. 

As discussed, the above TRA as three equivalent IA plans: broadcast based matrix multiplication (BMM), cross-product based matrix multiplication (CMM), and replication-based matrix multiplication (RMM). We compare these three IA implementations with Intel's version of ScaLAPACK \cite{choi1992scalapack} which realizes the classic SUMMA \cite{van1997summa}. We also compare with our own, hand-coded version of the classical 2.5D matrix multiply algorithm \cite{solomonik2011communication}, implemented on top of MPI \cite{barker2015message}; with Dask \cite{dask}, a popular distributed analytic tool with a Python interface \cite{rocklin2015dask}; and with PETSc \cite{petsc}, a popular high-performance distributed computing library \cite{balay2019petsc}.
All methods are benchmarked over Amazon EC2 clusters with 5, 10 or 15 \texttt{r5d.2xlarge} instances (each with 8 vCPU, 64 GB RAM, and connected by up to 10 Gb/s interconnect).
 Note that we have made reasonable amount of effort to tune the hyper-parameters in the alternative solutions (e.g., grid size, thread number, initial layout, etc.) and report the best results. 
Results are in Table \ref{table:mm}. In Table \ref{table:mm_cost}, we report the IA cost (as computed in Section \ref{sec:cost}) predicted for a 10-node cluster.

\subsection{Nearest Neighbor Search} 
We use TRA to implement a nearest neighbor search problem in a Riemannian metric space encoded by matrix $\textbf{A} \in \mathbb{R}^{D \times D}$, where given a query vector $\textbf{x}_q \in \mathbb{R}^{1 \times D}$ and a candidate set $\textbf{X} \in \mathbb{R}^{N \times D}$, the goal is to find the $i$-th row in the matrix that minimizes: 
$d_{\textbf{A}}\left(\textbf{x}_i, \textbf{x}_q\right) = \left(\textbf{x}_i - \textbf{x}_q\right)\textbf{A}\left(\textbf{x}_i - \textbf{x}_q\right)^T$. Suppose $\textbf{x}_q$, $\textbf{X}$, $\textbf{A}$ are stored in tensor relation $\texttt{R}_{\textbf{x}_q}$, $\texttt{R}_{\textbf{X}}$ and $\texttt{R}_{\textbf{A}}$, the corresponding TRA program can be encoded as:
{\small
\begin{align*}
    \texttt{R}_{\textbf{diff}} = &\Join_{\left(\langle 1 \rangle, \langle 1 \rangle, \texttt{matVecSub}\right)} \left( \texttt{R}_{\textbf{x}_q}, \texttt{R}_{\textbf{X}} \right) \\
    \texttt{R}_{\textbf{proj}} = &\Sigma_{\left(\langle 0, 2 \rangle, \texttt{matAdd}\right)} \left( \Join_{\left(\langle 1 \rangle, \langle 0 \rangle, \texttt{matMul}\right)} \left(\texttt{R}_{\textbf{diff}_{}} , \texttt{R}_{\textbf{A}}\right)\right) \\
    \texttt{R}_{\textbf{dist}} =&\lambda_{\left(\texttt{rowSum}\right)}\left(\Sigma_{\left(\langle 0 \rangle, \texttt{matAdd}\right)} \left( \Join_{\left(\langle 0,1 \rangle, \langle 0,1 \rangle, \texttt{elemMul}\right)} \left(\texttt{R}_{\textbf{proj}} , \texttt{R}_{\textbf{diff}}\right)\right)\right)\\
    \texttt{R}_{\textbf{min}} = &\Sigma_{\left(\langle \rangle, \texttt{minIndex}\right)}\left( \texttt{R}_{\textbf{dist}} \right).
\end{align*}
}%
\noindent where $\texttt{matVecSub}$ is matrix-vector subtraction, 
$\texttt{elemMul}$ is element-wise matrix multiplication (Hadamard product),
$\texttt{minIndex}$ finds the minimal element's index.
We hand-compile this into an expression in the IA, and then use the various equivalence rules to produce two different implementations: \texttt{Opt4Horizontal} and \texttt{Opt4Vertical}.
\texttt{Opt4Horizontal}
will broadcast $\texttt{R}_{\textbf{x}_q}$ and $\texttt{R}_{\textbf{A}}$ to each compute site and partition $\texttt{R}_{\textbf{X}}$ by dimension $0$; then the computation of $\texttt{R}_{\textbf{diff}} $, $\texttt{R}_{\textbf{proj}}$, and $\texttt{R}_{\textbf{dist}}$ will be conducted by local operations.  
\texttt{Opt4Vertical} will first broadcast $\texttt{R}_{\textbf{x}_q}$ to each site and compute $\texttt{R}_{\textbf{diff}}$, then partition $\texttt{R}_{\textbf{diff}}$ by dimension $1$ and partition $\texttt{R}_{\textbf{A}}$ by dimension $0$ so that $\texttt{R}_{\textbf{proj}}$ is computed in a cross-product based matrix multiplication. 
For the the \texttt{Opt4Horizontal} IA implementation, $\texttt{R}_{\textbf{x}_q}$, $\phys{R}_{\textbf{X}}$ and $\phys{R}_{\textbf{A}}$ are initially partitioned by dimension $0$. 
For the the \texttt{Opt4Vertical} IA implementation, $\phys{R}_{\textbf{x}_q}$, and $\phys{R}_{\textbf{A}}$ are initially partitioned by dimension $0$, while  $\phys{R}_{\textbf{X}}$ is initially partitioned by $1$.

We generate two data sets: (i) \texttt{Large}, with a large number of data points ({ $N=1.5\times10^5,1.5\times10^6$)} but small feature space ($D=6\times10^3$); and (ii) \texttt{Wide}, with a small number of data points ($N=6\times10^3$), with a large feature space ({ $D=3\times10^4,10^5$}). We execute this computation on compute clusters with 4, 8 or 12 \texttt{r5d.2xlarge} instances.

We also implemented the same computation using Dask\cite{dask}. And as a baseline, we compare the execution time with a PyTorch implementation that runs on a single site equipped with the same computing power as the TRA implementation: an \texttt{r5d.8xlarge} instance (with 32 vCPU, 256 GB RAM), an \texttt{r5d.16xlarge} instance (with 64 vCPU, 512 GB RAM) and an \texttt{r5d.24xlarge} instance (with 96 vCPU, 768 GB RAM).  Since the single-site implementation has zero communication overhead, this should be something of a lower-bound on the time required to run the computation. 
The results and predicted costs are enumerated in Table \ref{table:nns} and Table \ref{table:prednn}.

\subsection{Feed-Forward Neural Network} 

Lastly, we benchmark a training iteration of a two-layer FFNN for multiple label classification, computed over an input matrix.

Again, we compile the TRA program for FFNN learning by hand into the IA, and use the equivalence rules to produce two implementations. The first, called \texttt{TRA-DP}, resembles the classic data parallel implementation.
The second, called \texttt{TRA-MP}, corresponds to an intra-operation model parallel plan. 

We compare these two IA plans with the state-of-the-art data parallel implementation provided by PyTorch 1.7.1 \cite{li2020pytorch} and TensorFlow 2.4.1 \cite{abadi2016tensorflow}.  We also compare with the same computation written on top of Dask \cite{dask}, and hand-coded using ScaLAPACK \cite{choi1992scalapack}. Note that these two options do not fully support GPU.  

Two data sets are considered.  First, the data from the Google speech recognition task \cite{warden2018speech}, where a $1600$ feature vector is extracted from audio wave-forms; the goal is to identify 10 keywords ($D=1600$ and $L=10$); for this task, we train a very wide hidden layer with large number of neurons where $H = 1\times10^5$, $1.5\times10^5$, or $2\times10^5$; a batch size of $10^4$ ($N=10^4$) are used for min-batch SGD. Second, we consider the AmazonCat-14K \cite{mcauley2015image, mcauley2015inferring} benchmark, which is an extreme multi-label (XML) classification dataset including a large number of features ($D = 597540$) and labels ($L = 14588$); we train a relatively narrow network with $H = 0.5\times10^3, 1\times10^3$,  $3\times10^3$, $5\times10^3$, or $7\times10^3$; a batch size of $10^3$ ($N=10^3$) are used for mini-batch SGD.
Each is executed on CPU clusters with 2, 5 or 10 \texttt{r5dn.2xlarge} instances connected by up to 25 Gb/s interconnect) and GPU clusters with 2, 5 or 10 \texttt{p3.2xlarge} instances (each with a NVIDIA Tesla V100 GPU, and connected by 10 Gb/s interconnect).
The results for Google speech are listed in Table \ref{table:nn_speech}; for Amazon-XML in Table \ref{table:nn_xml}.  Predicted costs are given in Table \ref{table:ffnn_cost_pred}.

{
\begin{table}[]
\small
\begin{center}
\caption{SGD iteration time: FFNN for Google Speech.}

\begin{tabular}{|c||c|c|c||c|c|c|}
\hline
Cluster            & \multicolumn{3}{c||}{CPU} & \multicolumn{3}{c|}{GPU} \\ \hline
 Nodes & 2      & 5      & 10     & 2      & 5      & 10     \\ 
\hline
\multicolumn{7}{|c|}{\texttt{100k Neurons}} \\ 
\hline
PyTorch-DP  & 11.16s &  6.15s &  4.75s & 0.99s &  1.19s & 1.27s \\
TF-DP  &  11.93s &  7.32s &  5.51s &  0.87s &   1.13s &  1.17s  \\
ScaLAPCK  &  8.52s &  4.97s &  2.79s &  NA &   NA &  NA  \\
Dask  &  62.57s &  56.57s &  49.63s &  NA &   NA &  NA  \\
TRA-DP      & 11.62s &  6.51s &  5.20s & 1.49s &  1.59s & 1.63s \\
TRA-MP      & 26.56s & 28.71s & 29.09s & 7.01s & 11.56s & Fail \\
\hline

\multicolumn{7}{|c|}{\texttt{150k Neurons}} \\ 
\hline
PyTorch-DP  & 14.28s &  9.46s &  6.54s & 1.18s &  1.65s & 1.78s  \\
TF-DP  &  16.68s &  10.69s &  8.43s &  1.16s &   1.62s &  1.75s  \\
ScaLAPCK  &  13.45s &  7.48s &  3.87s &  NA &   NA &  NA  \\
Dask  &  96.56s &  85.32s &  77.07s &  NA &   NA &  NA  \\
TRA-DP      & 14.52s &  9.68s &  7.56s & 2.15s &  2.22s & 2.23s \\
TRA-MP      & 33.20s & 42.80s & 43.10s &  Fail &   Fail & Fail \\
\hline

\multicolumn{7}{|c|}{\texttt{200k Neurons}} \\ 
\hline
PyTorch-DP  & 17.25s & 11.94s &  9.30s & Fail &  2.09s & 2.42s \\
TF-DP  &  21.36s &  13.21s &  11.21s &  1.52s &   2.12s &  2.46s  \\
ScaLAPCK  &  17.18s &  10.05s &  5.06s &  NA &   NA &  NA  \\
Dask  &  136.66s &  112.72s &  104.01s &  NA &   NA &  NA  \\
TRA-DP      & 17.89s & 12.51s &  9.67s & 2.94s &  2.80s & 2.85s \\
TRA-MP      & 37.82s & 54.23s & 59.84s & Fail &   Fail & Fail \\
\hline

\end{tabular}
\label{table:nn_speech}
\end{center}
\end{table}
}

{
\begin{table}[]
\begin{center}
\small
\caption{SGD iteration time: FFNN for Amazon-XML.}
\begin{tabular}{|c||c|c|c||c|c|c|}
\hline
Cluster            & \multicolumn{3}{c||}{CPU} & \multicolumn{3}{c|}{GPU} \\ \hline
 Nodes & 2      & 5      & 10     & 2      & 5      & 10     \\ 
\hline
\multicolumn{7}{|c|}{ \texttt{0.5k Neurons}} \\ 
\hline
PyTorch-DP  &   3.58s &   4.51s  &   6.41s &  1.46s &  2.11s &  2.19s \\
TF-DP &  5.94s &  7.81s &  8.96s &  1.21s &  1.85s &  2.11s \\
ScaLAPCK  &  4.92s &  2.91s &  1.73s &  NA &   NA &  NA  \\
Dask  &  27.96s &  27.01s &  22.69s &  NA &   NA &  NA  \\
TRA-DP      &   4.77s &   5.13s  &   7.84s &  2.42s &  2.48s &  2.61s \\
TRA-MP      &   2.18s &   1.41s  &   0.83s &  0.15s &  0.12s &  0.09s \\
\hline

\multicolumn{7}{|c|}{\texttt{1k Neurons}} \\ 
\hline
PyTorch-DP  &  9.74s &  10.29s  &  10.34s & 2.67s & 3.76s & 4.20s \\
TF-DP &  Fail &  Fail &  Fail &  Fail &  Fail &  Fail \\
ScaLAPCK  &  8.16s &  6.65s &  2.47s &  NA &   NA &  NA  \\
Dask  &  45.40s &  42.15s &  29.34s &  NA &   NA &  NA  \\
TRA-DP      & 12.50s &  14.29s  &  15.68s & 4.67s & 4.69s & 4.73s \\
TRA-MP      & 3.86s  &   2.79s  &   1.70s & 0.40s & 0.37s & 0.35s \\
\hline

\multicolumn{7}{|c|}{\texttt{3k Neurons}} \\ 
\hline
PyTorch-DP  & 25.46s &  29.04s & 30.51s & Fail  &  Fail  &  Fail  \\
TF-DP &  Fail &  Fail &  Fail &  Fail &  Fail &  Fail \\
ScaLAPCK  &  17.56s &  9.59s &  7.91s &  NA &   NA &  NA  \\
Dask  &  103.83s &  89.09s &  81.56s &  NA &   NA &  NA  \\
TRA-DP      & 26.59s &  38.15s & 46.06s &  Fail & 12.74s & 13.13s \\
TRA-MP      & 10.57s &   6.36s &  3.88s &  Fail &  0.54s &  0.44s \\
\hline

\multicolumn{7}{|c|}{\texttt{5k Neurons}} \\ 
\hline
PyTorch-DP  & 34.05s & 46.53s & 50.17s & Fail & Fail  &  Fail \\
TF-DP &  Fail &  Fail &  Fail &  Fail &  Fail &  Fail \\
ScaLAPCK  &  23.21s &  11.65s &  8.33s &  NA &   NA &  NA  \\
 Dask  &  246.56s &  143.86s &  127.26s &  NA &   NA &  NA  \\
TRA-DP      & 44.12s & 68.54s & 75.15s & Fail & Fail  &  Fail \\
TRA-MP      & 18.59s &  8.07s &  5.75s & Fail & 0.59s & 0.48s \\
\hline

\multicolumn{7}{|c|}{\texttt{7k Neurons}} \\ 
\hline
PyTorch-DP  &  Fail  &  Fail  &   Fail  &  Fail &  Fail  &  Fail \\
TF-DP &  Fail &  Fail &  Fail &  Fail &  Fail &  Fail \\
ScaLAPCK  &  29.19s &  14.04s &  9.57s &  NA &   NA &  NA  \\
Dask  &  Fail &  Fail &  Fail &  NA &   NA &  NA  \\
TRA-DP      & 60.28s & 89.36s & 107.86s &  Fail &  Fail  &  Fail \\
TRA-MP      & 21.35s & 12.12s &  7.854s &  Fail &  Fail  & 0.73s \\
\hline

\end{tabular}

\label{table:nn_xml}
\end{center}
\end{table}
}

{
\begin{table}[t]
\small
\begin{center}
\caption{Predicted FFNN costs for a 5-node cluster.}
\begin{tabular}{|c||c|c|}
\hline
                      & TRA-DP          &        TRA-MP \\ 
\hline
\texttt{Google speech 100k}    &  $9.7 \times 10^{8}$    &   $1.0 \times 10^{10}$     \\              
\texttt{Google speech 150k}    &  $1.5 \times 10^{9}$    &   $1.5 \times 10^{10}$       \\
\texttt{Google speech 200k}    &  $1.9 \times 10^{9}$    &   $2.0 \times 10^{10}$  \\
\hline
\texttt{Amazon XML 1k}         &  $3.7 \times 10^{9}$    &   $1.0 \times 10^{7}$  \\
\texttt{Amazon XML 3k}         &  $1.1 \times 10^{10}$   &   $3.0 \times 10^{7}$  \\
\texttt{Amazon XML 5k}         &  $1.8 \times 10^{10}$   &   $5.0 \times 10^{7}$   \\
\texttt{Amazon XML 7k}         &  $2.6 \times 10^{10}$   &   $7.0 \times 10^{7}$   \\
\hline
\end{tabular}

\label{table:ffnn_cost_pred}
\end{center}
\end{table}
}

\subsection{Discussion}
\label{sec:exp_disc}

First, the experiments do seem to show that the TRA provides an abstraction upon which a variety of ML computations can be mapped. Further, the TRA seems to provide for good performance.  On the matrix multiplication experiments, the best TRA-based implementations were at least competitive with ScaLAPACK as well as our hand-coded MPI-based implementation (we observed 29s for the TRA-based CMM vs. 23s for hand-coded MPI in the ``two general marices'' case, 31s for the TRA-based BMM vs. 21s for hand-coded MPI in the ``two large dims case'') even beating them both (23s for the TAR-based CMM vs. 35s for hand-coded MPI) in the ``common large dim case''.  On the FFNN experiments, the best TRA-based implementation for each task was about 1.5$\times$ times as slow as the hand-constructed ScaLAPACK implementation for the Google data set, but considerably faster than ScaLAPACK for the more challenging Amazon data set. In general it is fair to say that the best TRA-based implementation for each task was at least competitive with the ScaLAPACK and MPI-based codes.  The fact that there is not a significant performance hit moving from a special-purpose tool requiring significant programmer expertise to a general-purpose implementation abstraction seems to argue that in fact, a TRA-based back-end can provide state-of-the-art performance.

It is also instructive to compare our TRA-based implementations with the other, more user-friendly tools tested, which in practice would be more reasonable alternatives to an ML system with a TRA-based back-end. Dask was not competitive, and was often one or two orders of magnitude slower.  On the FFNN experiments, PyTorch was generally better performing than TensorFlow in CPU clusters, while both systems perform almost identically in GPU clusters. For Google speech, the optimal partition schema is identical to data parallelism, where the best TRA-based option is able to closely match PyTorch's speed. Further, while PyTorch failed on the larger Google computations in a 2-GPU cluster, the TRA implementation was able to run to completion. On the even larger, extreme classification problem, the TRA-MP (model parallel) IA was much, much faster than PyTorch, (where TensorFlow fails in most cases since it does not allow a parameter matrix to exceed 2 GB), and much more scalable. PyTorch also cannot handle the huge matrices required to power this computation in some settings. 

The final question we wanted to address was whether the TRA is amenable to automatic optimization.
Note that in each case, there was one IA implementation that was suitable for the input data, and one that was not; the difference between the two was often significant.  In a system based upon the TRA, it would be crucial to automatically choose the suitable implementation.  We found that in each case, the simple cost model from Section \ref{sec:cost} would have chosen the correct implementation.  For example, consider Table \ref{table:ffnn_cost_pred}.  In each case, the cost metric correctly assigns the lower cost to the appropriate IA computation: TRA-DP for the smaller, Google problem, and TRA-MP for the larger, extreme classification problem.  These results suggest that it should easily be possible to perform cost-based optimization over the IA.

\section{Related Work}
\label{sec_rel}


\noindent  Our focus has been on the proper implementation abstraction for ML systems. The TRA is ``front-end agnostic.'' Still, there has been considerable interest in programming and compilation for such systems. FAQ, by Khamis et al. \cite{abo2016faq}, considers how to compute Einstein-notation-like expressions over semi-rings. Effectively, FAQ re-writes such expressions so that they can easily be computed using the ``OutsideIn'' method for first determining the non-zero entries using a series of joins, followed by the computation of the values. Laue et al. \cite{laue2020simple} propose a variant on the Ricci calculus for computing tensor-based derivatives using the Einstein notation. Tensor Comprehensions are an Einstein-like programming language and associated compiler that is able to produce efficient CUDA kernels \cite{vasilache2018tensor}; the tensor algebra compiler is a similar effort \cite{kjolstad2017tensor}. Our efforts are complementary.  One could imagine, for example, using FAQ-like algorithms along with a compiler for high-performance kernels to generate expressions for a TRA-based back-end.

Classic data-flow systems have been modified to support distributed machine learning.
Both Spark \cite{zaharia2010spark} and SystemML \cite{ghoting2011systemml} provide native libraries for deep learning. 
A set of deep learning frameworks can run on top of Spark, such as TensorFrames \cite{hunter2016tensorframes}, Deeplearning4j \cite{team2016deeplearning4j}, SparkNet\cite{moritz2015sparknet} and BigDL \cite{dai2019bigdl}. Conceptually, these deep learning frameworks are related to the TRA as they allow the distribution of ML computations.  Consider TensorFrames. TensorFrames allows the items in a Spark DataFrame to be operated on by a TensorFlow computation. One could view those TensorFlow computations as being similar to the kernels applied by TRA, and the Spark operations used to manipulate the data as being similar to the the joins, aggregations, and so on offered by the TRA. The key difference is that while these systems are each significant engineering efforts aimed at marrying different technologies (TensorFlow and Spark in the case of TensorFrames), the TRA is designed as a generic back-end.  In fact, a TensorFrames-like programming model could easily be mapped onto TRA, with \texttt{mapRows}, \texttt{aggregate}, etc., being mapped to the appropriate TRA operations, and the TensorFlow computations run as kernels.



Relational systems have long been proposed for ML. 
MLog \cite{li2017mlog} is a declarative relational system managing data movement, data persistency, and training batch generation. Similar ideas have been applied in  
\cite{abo2018database} for feature extraction queries over multi-relation databases,
and \cite{khamis2018ac} for optimizing sparse tensor computations constructed from relational tables. 
Recently, relational systems have also been considered as runtime engine (instead of an efficient data loader) for distributed ML.
DB4ML \cite{jasny2020db4ml} proposes user-defined iterative transactions.
Multi-dimensional-recursion has been built on top of SimSQL \cite{cai2013simulation}, a distributed analytic database system, that can support neural network training \cite{jankov2019declarative}. 
The idea of moving past relations onto arrays as a database data model, is long-standing (e.g., consider Baumann's work on Rasdaman \cite{baumann1998multidimensional}). 
SciDB \cite{brown2010overview} is a well-known system following this idea.
LevelHeaded \cite{aberger2018levelheaded} uses a special key-value structure to support linear operations.
MATLANG \cite{barcelo2019expressiveness} introduces a language for matrix manipulation.
TensorDB \cite{kim2014tensordb,kim2014efficient} is a database system that can perform tensor manipulation.
LARA\cite{hutchison2017laradb} proposes an algebra with tuple-wise operators, attribute-wise operators, and tuple extensions, then defines linear and relational algebra operations using these primitives.
RMA \cite{dolmatova2020relational} attempts to bridge the gap between relations and matrices.
While related, these systems attempt to implement tensor computations as algebraic expressions (e.g., a join followed by an aggregation) over relations of \texttt{(key, value)} pairs. This requires pushing a huge number of pairs through the system, which introduces significant overhead.


\vspace{-10pt}

\section{Conclusion}

We have introduced the tensor relational algebra (TRA), and suggested this as the interface that could be exported by the back-end of a machine learning system.  
We have showed through extensive experimentation that a computation expressed in the TRA then transformed into the implementation algebra and  optimized, is competitive with (and often faster than) other options, including HPC softwares such as ScaLAPACK, and ML softwares such as TensorFlow and PyTorch.

There are many avenues for future work. TRA is not meant to be a user-facing programming language. Thus, a key question is: can a language such as Tensor Comprehensions or Einstein notation be compiled into TRA? At a high level, this should not be too difficult, as these languages match indices in different tensors (which is easily implemented as a join) and then sum out dimensions (aggregation). But there are many details to consider. The TRA uses arrays or ``chunks'' for speed.  How to automatically block or chunk a tensor computation? How to automatically generate the compute kernels? Sparsity is also an important issue.  A compiler could also decide to store a sparse tensor using arrays that do not have zero dimensions, but where those arrays are stored sparsely, with a high-performance kernel generated to handle the specific sparsity pattern.

\appendix

\begin{acks}
This work was supported by an NIH CTSA, award no. UL1TR003167 and by the NSF under grant nos. 1918651, 1910803, 2008240 and 1842494. We also thank the anonymous reviewers for their insightful comments on earlier versions of the paper.
\end{acks}


\balance
\bibliographystyle{ACM-Reference-Format}
\bibliography{tra}

\end{document}